\documentclass[12pt,letterpaper]{article}

\usepackage[body={18cm, 25cm}]{geometry}
\usepackage[utf8]{inputenc}
\usepackage{paralist}
\usepackage{multicol}
\usepackage{multirow}
\usepackage{float}
\usepackage{amsmath}
\usepackage{amsfonts}
\usepackage{amssymb}
\usepackage{graphicx}
\usepackage{geometry}
\usepackage{color}
\usepackage[]{algorithm2e}

\usepackage[square,comma,numbers,sort&compress,super]{natbib}
\usepackage{subfig}
\usepackage[pdftex, pdftitle={}, pdfauthor={}, pdfsubject={}, pdfkeywords={}, pdfpagemode=UseOutlines,bookmarks,bookmarksopen,pdfstartview=FitH,colorlinks,linkcolor=blue, urlcolor=black, citecolor=blue]{hyperref} 
\usepackage[export]{adjustbox}
\usepackage{tabularx}
\usepackage{tabulary}
\usepackage[table]{xcolor}
\usepackage{caption}
\usepackage[noabbrev]{cleveref}
\usepackage{indentfirst}
\usepackage{authblk}
\usepackage[utf8]{inputenc}
\usepackage[english]{babel}
\usepackage{color}

\usepackage[symbol]{footmisc}

\title{\textbf{Computational Characterization of the Wave Propagation Behaviour of Multi-Stable  Periodic Cellular Materials\footnote{Preprint submitted to Extreme Mechanics Letters}}}
\author[1]{\textit{Camilo Valencia}}
\author[2,4]{\textit{David Restrepo}}
\author[3]{\textit{Nilesh D. Mankame}}
\author[2\thanks{Corresponding author: zavattie@purdue.edu}]{\textit{Pablo D. Zavattieri}}
\author[1\thanks{Corresponding author: jgomezc1@eafit.edu.co}]{\textit{Juan Gomez}}
\affil[1]{Civil Engineering Department, Universidad EAFIT, Medellín, 050022, Colombia}
\affil[2]{Lyles School of Civil Engineering, Purdue University, West Lafayette, IN 47907, USA}
\affil[3]{Vehicle Systems Research Laboratory,  General Motors Global Research \& Development, Warren, MI 48092, USA.}   
\affil[4]{Department of Mechanical Engineering, The University of Texas at San Antonio, San Antonio, TX 78249, USA}   
\date{}

\graphicspath{{./figs/}}

\begin{document}

\maketitle

\begin{abstract}
  In this work, we present a computational analysis of the planar wave propagation behavior of a one-dimensional periodic multi-stable cellular material. Wave propagation in these materials is interesting because they combine the ability of periodic cellular materials to exhibit stop and pass bands with the ability to dissipate energy through cell-level elastic instabilities. Here, we use Bloch periodic boundary conditions  to compute the dispersion curves and introduce a new approach for computing wide band directionality plots. Also, we deconstruct the wave propagation behavior of this material to identify the contributions from its various structural elements by progressively building the unit cell, structural element by element, from a simple, homogeneous, isotropic primitive. Direct integration time domain analyses of a representative volume element  at a few salient frequencies in the stop and pass bands are used to confirm the existence of partial band gaps in the response of the cellular material. Insights gained from the above analyses are then used to explore modifications of the unit cell that allow the user to tune the band gaps in the response of the material.
  We show that this material behaves like a locally resonant material that exhibits low frequency band gaps for small amplitude planar waves. Moreover,  modulating the geometry or material of the central bar in the unit cell provides a path to adjust the position of the band gaps in the material response. Also, our results show that the material exhibits highly anisotropic wave propagation behavior that stems from the anisotropy in its mechanical structure. Notably, we found that unlike other multi-stable cellular materials reported in the literature, in the system studied in this work, the configurational changes in the unit cell corresponding to its different stable phases do not significantly alter the wave propagation behavior of the material. 
\end{abstract}

\section{Introduction}
Metamaterials are a new class of materials that derive their unique functionality from their cellular structure rather than the chemical composition of the material from which they are made. These materials offer the promise of exciting new applications because they exhibit properties that are not found in naturally occuring bulk materials. Some examples include a negative refraction index material which can be used in flat loss-less lenses \cite{venema2002negative},  room temperature THz modulators for ultrafast wireless communication \cite{chen2006active},  and acoustic negative refraction for high resolution medical imaging \cite{lu2009phononic}. Metastructures share the periodic cellular construction and hence, many of the unique functional attributes of metamaterials. Metastructures have been designed to function as  compact seismic mitigators that can mitigate the impact of earthquakes on sensitive buildings \cite{krodel2015wide} and as reusable solid-state energy dissipators for seismic mitigation and sports helmets \citep{restrepo2015phase,correa2015negative}. Metamaterials (or metastructures) that are designed to control the wave propagation behavior of elastic or acoustic waves are called phononic metamaterials (or metastructures). Kadic et al. \cite{kadic2013metamaterials} and Hussein et al. \cite{hussein2014dynamics} provide excellent reviews of the state of the art in metamaterials at large and phononic metamaterials respectively. 

Phase transforming cellular materials (PXCMs) are periodic cellular materials whose unit cells exhibit a snap-through instability that arises from the traversal of an elastic limit point in its force - displacement response \cite{restrepo2015phase}. When a cell in the material 'snaps through' under external load, some of the strain energy stored in the material due to the external work done up to that point is released as kinetic energy. This gives rise to oscillatory waves that dissipate the released energy, which eventually ends up as heat. This is similar to the mechanics of the 'waiting link' structure of Cherakev et al. \cite{cherkaev2005transition}.  However, unlike the waiting link structure,  if the unit cells in PXCMs are sized properly the material can remain in the elastic regime during the loading and unloading steps. Thus, PXCMs offer a way to realize reusable solid state energy dissipators. The quasi-static behavior of these materials in response to cyclic loads \citep{restrepo2015phase,che2017three}, high strain rate monotonic loads \cite{liu2018dynamic}, and impact loads \citep{correa2015negative,debeau2018impact} has been studied via analytical, computational and experimental means in recent years.

In this paper we study the propagation of small amplitude planar waves in a one dimensional PXCM with bistable unit cells (see figure 1) similar to the PXCM studied in Restrepo et al. \citep{restrepo2015phase}. 

Although the PXCM is functionally one dimensional (i.e. it dissipates significant amounts of energy for only for loads applied along one material direction), it is structurally a two dimensional material that is periodic in two orthogonal directions. Structural periodicity gives rise to peculiar wave propagation behaviors in periodic media such as the presence of band gaps, which are frequency intervals in which wave propagation is highly attenuated, and anisotropic energy propagation. The topology (i.e. material connectivity) of a unit cell of the PXCM material does not change when it transitions from its undeformed stable configuration to the deformed stable configuration (see figure 1c). However, this transformation is associated with significant changes in the relative density, periodicity in the Y direction, and the shape of the bent beams in the cell. Prior work on planar wave propagation in two dimensional periodic cellular materials with low relative density has shown that all of these attributes - the shape of the cell edges \citep{gonella2008analysis, trainiti2016wave, casadei2013anisotropy}, the relative density \citep{gonella2008analysis, casadei2013anisotropy} and periodicity \citep{bertoldi2008wave} - play an important role in the linear wave propagation behavior of periodic cellular materials.

Changes in unit cell geometry such as the internal angles and edge thicknesses \citep{gonella2008analysis, casadei2013anisotropy}, modulated undulations in the cell edges \cite{trainiti2016wave} and hole shapes \citep{bertoldi2008wave} have been shown to result in dramatic changes in the band gap structure as well as the directionality of energy flow in two dimensional periodic cellular materials. However, once the material is fabricated its wave propagation behavior is fixed. In recent years there has been significant interest in periodic materials whose wave propagation behavior can be significantly altered after the material has been fabricated. Schaeffler and Ruzzene \cite{schaeffer2015wave} showed that switching between the various stable structural configurations in a magnetoelastic system could lead to large changes in the geometry of the unit cell, which in turn lead to significant changes in the wave propagation behavior. Large geometric changes in a unit cell are triggered by local (i.e. in every cell) instabilities in a periodic elastomeric material subjected to large global compressive strains. Bertoldi and Boyce \cite{bertoldi2008wave} exploited these changes and the concomittant redefinition of the periodicity of the material to induce large changes in the band gap structure of the material. Moreover, since these changes are achieved by elastic deformation in an elastomeric material, they can be reversed and the whole process can be repeated. Bernard and Mann \cite{bernard2013passive} showed that a periodic multistable material could be designed with very different behavior in its various stable states such that a bifurcation induced switching of stable states in this material could change its behavior from freely allowing wave propagation at a particular frequency to blocking propagation at that frequency. Meaud and Che \cite{meaud2017tuning} studied linear wave propagation in a one dimensional chain of bistable unit cells that are similar to the cells in the PXCM that is the focus of this work.  The stability of the unit cell changes from being bistable  for low values of the thickness of the inclined beams in the cell to being monostable for thicknesses higher than a critical value. Meaud and Che \cite{meaud2017tuning} noted that the stiffness of the unit cell in the deformed stable configuration approached zero as the thickness of the inclined beams approached the critical value. A dramatic difference in the wave propagation behaviors of the two stable configurations can be obtained if the unit cell was designed with the beam thickness close to its critical value.

Obtaining wide band low frequency band gaps is an important and practical challenge in the design of phononic metamaterials. The two common mechanisms that give rise to bandgaps in periodic media - Bragg scattering and local resonances - have been shown to result in structures that are impractical because they are too big, too heavy or have a narrow bandgap \citep{yilmaz2007phononic,matlack2016composite}. Researchers have tried to overcome these limitations by using arrays of local resonators that are tuned to different frequencies spread over a desired band gap interval \citep{krodel2015wide}, creating structures that couple a Bragg scattering bandgap and a nearby resonant bandgap to create a coupled bandgap \citep{yuan2013coupling, krushynska2017coupling}, using compliant displacement amplifiers embedded in a unit cell to boost the dynamic inertia properties of the cell \citep{yilmaz2007phononic} and coupling structural vibration modes with local resonator modes \citep{matlack2016composite}. All of these approaches have shown promise in creating wide band low frequency band gaps whose location and size can be adjusted by tuning the mass and stiffness distribution of the unit cell.

We build on the above work to show that the PXCM studied in this work can be designed to create low frequency band gaps for small amplitude planar waves.  We use Bloch analyses of unit cell and transient simulations of a representative volume element in a plane strain finite element framework  to study the dispersion (i.e. band gap) behavior and the directionality of vibratory energy flow in the material. We develop a new method to compute the directionality of energy flow in these materials by including contributions from several modes instead of limiting this to just the first few modes as is done typically. This allowed us to obtain a complete picture of energy flow in these materials over a wide band of frequencies. In order to  get a better understanding of the material response we progressively build the unit cell, structural element by element, starting from a homogeneous isotropic primitive. The response of the cell at each step along the way is determined. This helped us to deconstruct the response of the unit cell and understand how various structural elements of the cell contribute to its overall behavior. Single-frequency wavelets with characteristic frequencies inside and outside of the band-gaps in the dispersion diagram are applied to a representative volume element in a direct integration transient analysis to verify the wave propagation behavior predicted by the dispersion diagram.  This leads to several insights into the material response such as  (i) mechanisms responsible for the presence (or absence) of band gaps in vertical direction; (ii) structural elements that control the  location and bandwidth of the band gaps in vertical direction; and (iii) structural elements that control the directionality of vibratory energy propagation in the material.  These insights lead to the ability to tune the design of the unit cell for a desired band gap location.

The rest of the paper is organized in five sections. Section 2 summarizes background information regarding the basic mechanics of the PXCM material studied here and the Bloch analysis used in this study.  In Section 3 we describe a new method to compute the directionality of energy propagation in this material. Section 4 is the main section of this paper. Its various subsections cover the deconstruction of the overall wave propagation behavior of the PXCM material, transient simulations to verify the dispersive behavior at a few salient frequencies and a way to modulate the dispersive behavior by changing properties of the unit cell.  The last section discusses the differences between the wave propagation behaviors of the material studied here and similar materials reported in the literature. Strategies to widen the low frequency band gap are also discussed in this section.

\section{Background and theoretical aspects}
\subsection{Mechanics of the phase transforming cellular material}
\label{sec:PXCM}
\begin{figure} [H]
\centering
\subfloat{\includegraphics[width=0.4\textwidth]{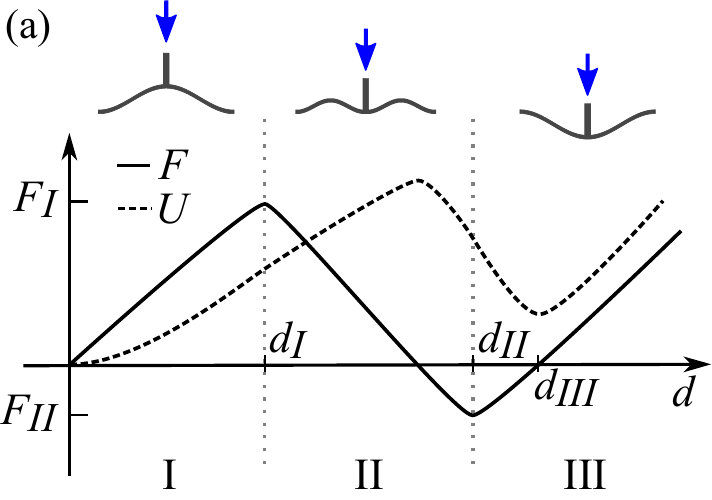}} \quad
\subfloat{\includegraphics[width=0.5\textwidth]{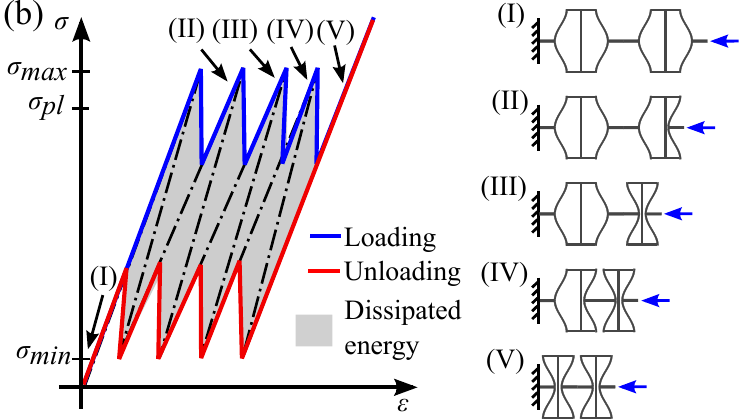}} \\
\subfloat{\includegraphics[width=0.54\textwidth]{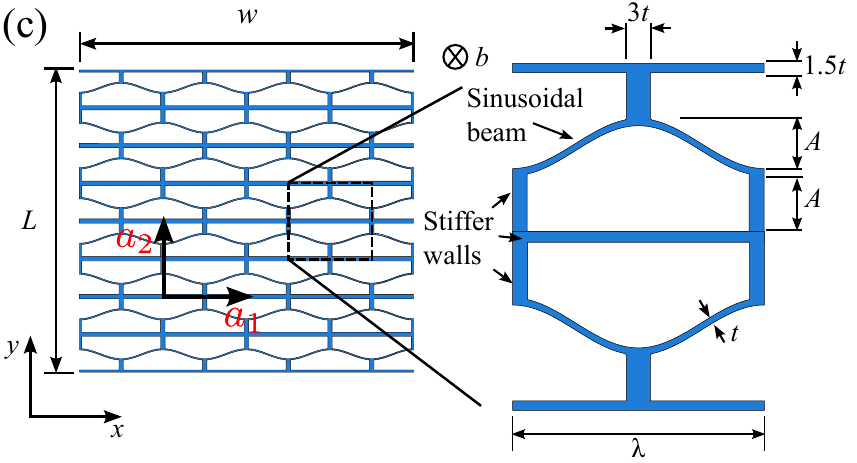}}
\subfloat{\includegraphics[width=0.45\textwidth]{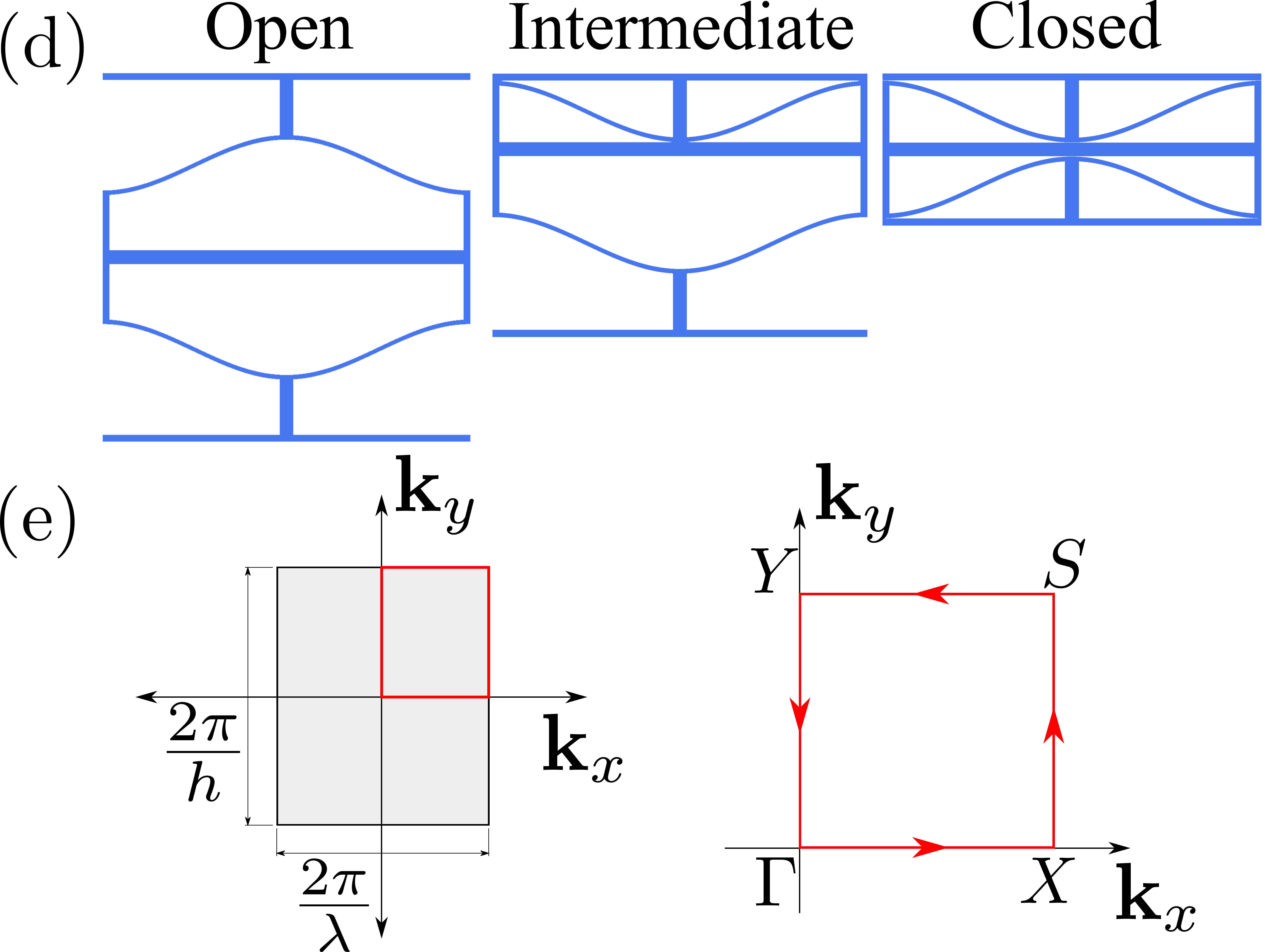}}\\
\caption{Phase Transforming Cellular Materials (PXCM).  (a) Schematic representation of the force-displacement ($F-d$) behavior and change of potential energy ($U$) as function of displacement for a bistable mechanism, (b) force-displacement behavior for a two unit cell PXCM sample, (c) geometric parameters for a PXCM sample and a constituent mechanism and (d) stable phases at the unit cell level. (e) First Brollouin zone and Irreducible Brillouin zone. Parameter $h$ is the total height of the unit cell at hand.}
\label{fig:PXCM1}
\end{figure}

A unit cells of a PXCM comprises two compliant bistable mechanisms. Each of these bistable mechanisms exhibits a force-displacement response that is characterized by two limit points ($(d_{I},F_{I})$ and $(d_{II},F_{II})$) as seen in \textbf{\cref{fig:PXCM1}}a.  This allows us to segment the response into  three characteristic regimes: regimes I and III that are characterized by a positive stiffness as they represent the deformation of stable configurations of the unit cell, and regime II that is characterized by a negative stiffness\cite{howell2001compliant,howell1994reliability}. These unit cells are arranged in a space filling  array to make  PXCMs. The mechanical response of a two unit cell material is as shown in \cref{fig:PXCM1}(b).

The unit cell configurations corressponding to the two stable equilibria are interpreted as phases of the PXCM, and a change in the configuration of its unit cell between stable configurations is interpreted as a phase transformation \cite{restrepo2015phase}. The propagation of the phase transformation corresponds to a progressive change of configurations from one row of bistable mechanisms to the next one,  leading to a serrated plateau in the force-dispalcement response.  Traversal of the limit points results in the non-equilibrium release of stored strain energy.  Thus, PXCMs exhibit hysteresis under mechanical loading and  the force-displacement has two finitely separated serrated plateaus that correspond to loading and unloading respectively.  is different for loading and unloading; therefore, PXCMs exhibit large hysteresis. If the unit cells are designed such that despite undergoing large deformations the strains in the cell elements do not exceed the elastic limit of the constituent material, there is no irreversible deformation during the loading and unloading of these materials. Hence,  the material can be used to dissipate energy multiple times.

In addition to energy dissipation, phase transformations in PXCMs are accompanied by large configurational changes in the unit cell. These changes affect macroscopic properties of the material such as its effective density. In this work we investigate the effect of these changes on the wave propagation behavior  of the material e.g. changes in the propagation of elastic waves over certain frequency bands (band-gaps), and changes in the direction of vibrational energy propagation through the material. 

In this work  we focus our analysis on the PXCM shown in \cref{fig:PXCM1}c which was introduced in \cite{restrepo2015phase}.  The unit cells of this PXCM are composed of two curved beams that are connected by stiffening walls, which provide local support and prevent transverse displacements. This lateral constraint allows the curved beams to function as bistable mechanisms. The shape of the curved beams is described by $y= (\frac{A}{2})[1-\sin(2\pi(x-\lambda/4)/\lambda)] $, where $\lambda$ is the wavelength; and $A$  the amplitude (peak to valley). Other geometric parameters for the unit cell are the  beam thickness $t$, out of plane depth $b$ and the thickness of the stiffening walls which is set to $1.5t$. The mechanical response of the curved beams is primarily determined by the amplitude to thickness ratio $Q=A/t$; bistable behavior is  obtained when $Q\geq2.41$ \cite{vangbo1998analytical,qiu2004curved,restrepo2015phase}. The maximum strain in the beams, $\varepsilon_{max}=2\pi^{2}\frac{tA}{\lambda^{2}}$, should be kept below the yield strain of the base material to avoid irreversible deformation and allow the material to be used multiple times.   In this work the  geometric parameters describing the unit cell of the PXCM were kept constant with values corresponding to $A = 9.04$ mm, $\lambda = 60$ mm, $b= 20.576$ mm, and $t = 0.742$ mm. 

Although this PXCM is a 2D cellular material, phase transformations take place only for loads in the $y$-direction and its deformation is mainly governed by the deflection of the curved beams. Three stable configurations can be identified at the unit cell level: open, intermediate, and closed as shown in \cref{fig:PXCM1}d.  
 
\subsection{Bloch's theorem}
\label{sec:theoback}

Spatially periodic materials like  PXCMs can be divided into identical finite-sized fundamental (unit) cells which fill out space after properly imposed tessellation operations.  For a wave propagating without attenuation through such a material  with  spatial periodicity $a$, Bloch's theorem \citep{brillouin2003wave} states that the local (i.e. over a unit cell) change in wave amplitude does not depend on the specific location of the cell in the periodic ensemble.  The displacement field over a unit cell satisfies: 

\begin{equation}
U(x+a) = U(x)e^{i\kappa \cdot a}
\label{eq:bloch}
\end{equation}

where $U(x)$ and $U(x+a)$ describe the response at the field points $x$ and $x+a$ respectively, $\kappa$ is the wave vector,  $a = a_1n_1 + a_2n_2 + a_3n_3$ is the lattice translation vector, $n_i$ are integers representing the number of translations along the $a_i$ direction, and the term $e^{i\kappa \cdot a}$ represents the change in the displacement field over the unit cell.  The Bloch theorem makes it  possible to characterize the wave propagation behavior of a periodic cellular material  by analyzing the wave motion only in its fundamental unit cell.


A unit cell of the PXCM in the open configuration is shown in  \cref{fig:PXCM1}(c). The primitive lattice vectors $a_1$ and $a_2$ defining the lattice periodicity are overlaid in black solid arrows.  We define a reciprocal lattice in the wave vector space ($\kappa=(\kappa_x,\kappa_y)$), and compute the dispersion relation ($\omega=\omega(\kappa_x,\kappa_y)$) in the reciprocal space by varying $\kappa$ along the first Brillouin Zone. The first Brillouin zone for this material can be further reduced to the Irreducible Brillouin Zone (IBZ) by accounting for the symmetries of the first Brillouin zone.  The irreducible Brillouin zone  corresponds to the smallest area over which we can vary $\kappa$  and still obtain a full representation of the dispersion relation. The first and the irreducible Brillouin zones for the PXCM are shown in \cref{fig:PXCM1}(e).

The dispersion relation of a material under free wave motion can be obtained by discretizing its equation of motion in a finite element framework.  The discretized equation of motion neglecting attenuation effects has the form:

\begin{equation}
\left[ K - \omega^2 M \right]U = 0 
\label{eq:discrete wave}
\end{equation}
\noindent

where $\omega$ is the frequency of harmonic wave propagation, $K$ and $M$ are respectively the stiffness and mass matrices for the unit cell, and $U$ is the nodal displacement vector.  Application of Bloch's Theorem \cref{eq:bloch} results in the generalized eigenvalue problem

\begin{equation}
\left[ K_R - \omega^2 M_R \right]U_R = 0 
\label{eq:eig}
\end{equation}
\noindent

where $K_R$, $M_R$ and $U_R$ are the reduced versions of $K$, $M$ and $U$ resulting after removing redundant equations from \cref{eq:discrete wave}.  The boundary of the IBZ is discretized and this eigenvalue problem is solved for a series of assigned values of the wave vector $\kappa = \langle \kappa_x,\kappa_y \rangle$ covering the boundary of the IBZ. Each assignment to a wave vector $\kappa$ yields a particular instance of the generalized eigenvalue problem given in \cref{eq:eig} and represents a plane wave propagating at frequency $\omega$. The corresponding solution gives all frequencies $\omega$  at which propagation is possible for that wave vector.  This leads to the dispersion relation $\omega=\omega(\kappa_x,\kappa_y)$.

Dispersion relations provide essential information about the wave propagation behavior in a material. They can be used to identify the speed at which energy is transported by the wave (wave group velocity), the directions along which the vibrational energy of the wave is propagating (directionality), and intervals of frequency where the material cannot propagate vibrational energy (frequency band-gaps). Dispersion relations are therefore fundamental representations of the wave propagation behavior of  periodic materials and their ability to propagate or attenuate waves at certain frequencies and along certain directions.

\subsection{Wide band energy propagation directionality}

The dispersion relations define a set of surfaces known as phase constant surfaces over the first Brillouin zone in the wave vector space. Each surface corresponds to a dispersion mode ($M_i$), and the number of these surfaces corresponds to the size of the eigenvalue problem in \cref{eq:eig}. These surfaces can be represented in two-dimensions as iso-frequency contour plots. The normal to any point on an iso-frequency contour corresponds to the direction of propagation of the energy (i.e. the direction of the group velocity) at that point and the gradient of the contour at the point yields the magnitude of the group velocity.  

The directionality of energy propagation can be inferred in many ways. Phani et al. \cite{phani2006wave} use direct inspection of the iso-frequency contours to infer directionality of energy propagation. Isotropic energy propagation results in circular contours at a given frequency, while anisotropic (or noticeably directional) energy propagation gives rise to lobes or spikes along preferred directions. Ruzzene et al. \cite{ruzzene2003wave} indicate directionality by marking the range over which the direction of the group velocity vector varies as we traverse a given iso-frequency contour. If the range spans a complete circle, it indicates isotropy in energy propagation at that frequency. Others (e.g. \cite{casadei2013anisotropy}) plot the vertical (Y) component of the group velocity against its horizontal (X) component for each point along a given frequency contour. Isotropic energy propagation leads to circles while anisotropic energy propagation gives rise to lobed traces in these plots. 

When the directionality of energy propagation for an entire mode that spans a frequency range is desired, a polar histogram can be constructed based on the group velocity field for that mode. In this work, we extend this notion by extending the range of frequencies that is represented in the directionality plots to cover all frequencies for which the chosen finite element discretization provides an accurate dynamic response.   

We propose a strategy to obtain wave propagation directionality in which the contributions of multiple modes Mi are considered. This approach allows a more complete description of the directional response in a material, and is valid over the entire frequency range of interest. We define the wave propagation directionality, $D$, as

\begin{equation}
D = \sum_{\substack{i\\e>tol}}d_i(\theta) \quad ; \quad d_i(\theta) = C\left( \frac{\nabla M_i}{\alpha} \right)
\label{eq:dir}
\end{equation}

where $\nabla M_i$ is the gradient of the $i$-th mode in the dispersion relation, $\alpha$ is the P-wave speed of the base material, $tol$ is a predefined tolerance, $\theta = [0,2\pi]$ is the angle defining the propagation direction. 

\begin{figure}[H]
\centering
\subfloat{\includegraphics[width=0.9\textwidth]{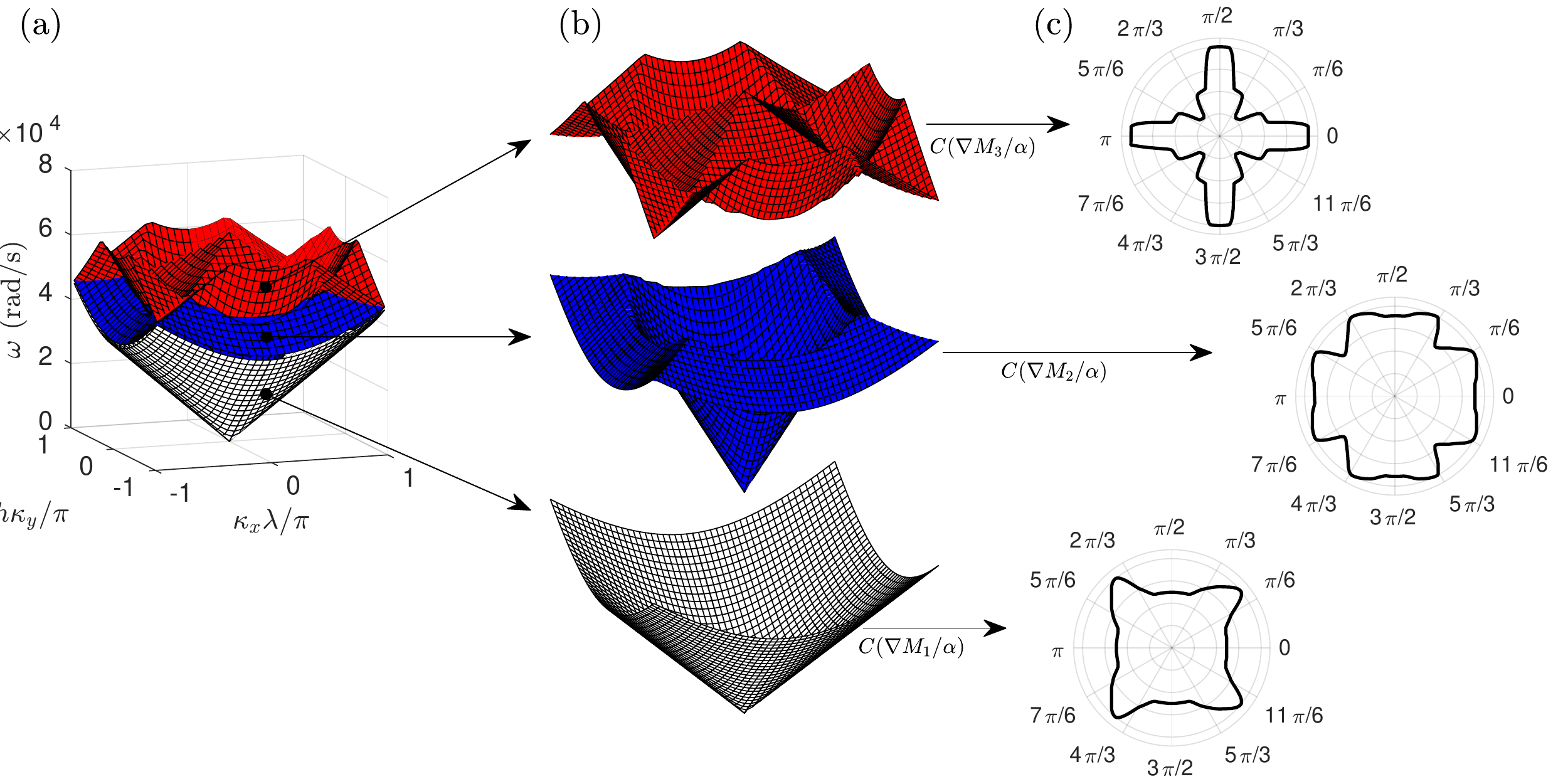}}\\
\subfloat{\includegraphics[width=0.7\textwidth]{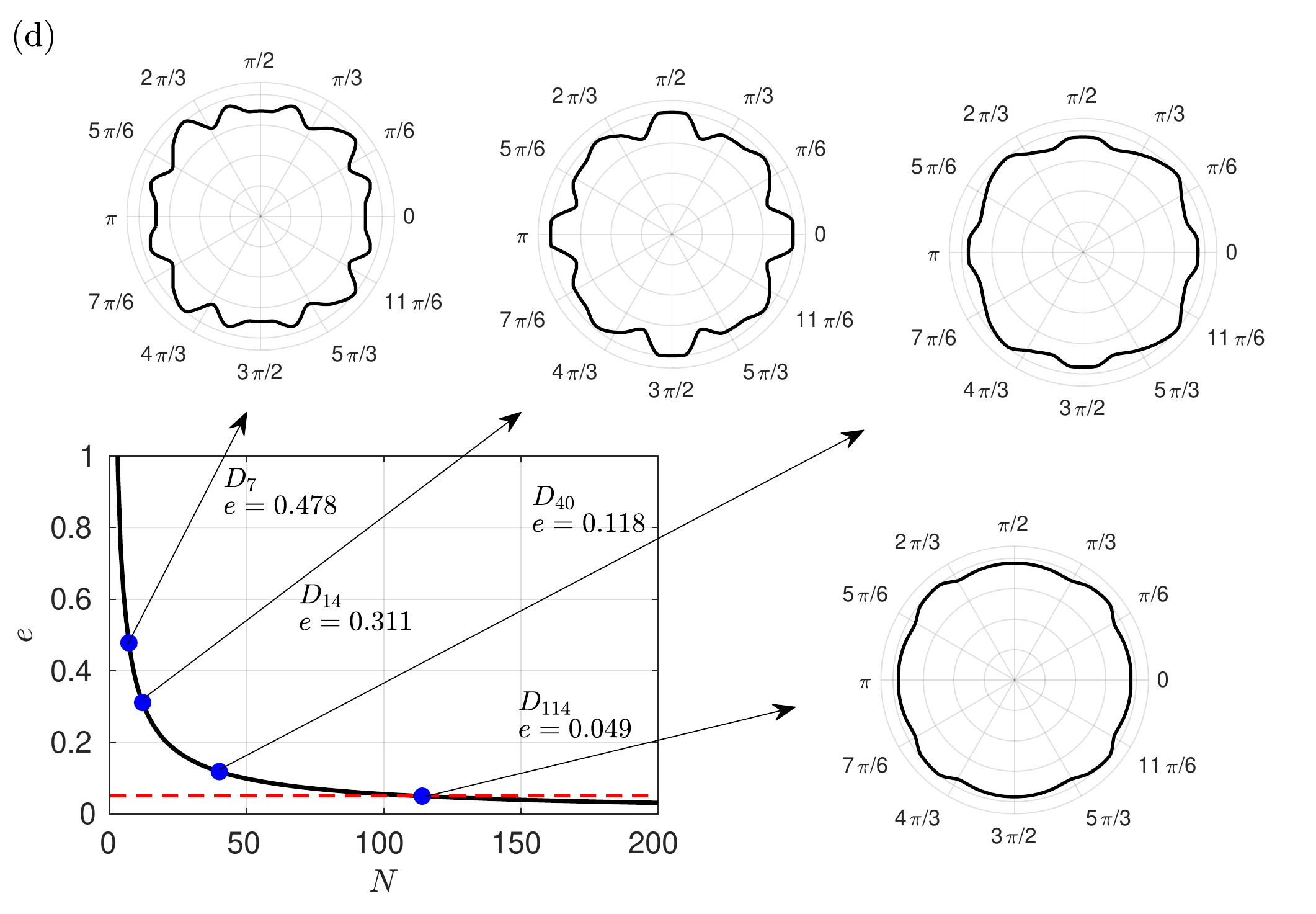}}
\caption{Result from appliying \cref{eq:dir} to a homogeneous material unit cell with mechanical properties $E = 1$e9Pa, $\nu = 0.3$ and $\rho = 1$g/cm$^3$. (a)-(b) Phase constant surfaces for the first three modes of the unit cell. Surfaces white, blue and red correspond to modes $M_1$, $M_2$ and $M_3$ respectively. (c) Polar representation of $d_1$, $d_2$ and $d_3$; that are obtained by applying the $C$-operator to modes $M_1$, $M_2$ and $M_3$ respectively. (d) Evolution of directionality $D$ and error $e$. The term $D_N$ is used to denote the directionality plot after considering $N$ modes. The predefined tolerance $tol = 0.05$ is reached after 114 modes in this case.\label{fig:modes}}
\end{figure}

In this definition, operator $C(V)$ performs the following operations over every vector $v_t$ in the vector field $V$: (i) Calculates the direction and magnitude of $v_t$. (ii) Normalizes the magnitude of $v_t$ by $\alpha$. (iii) Accumulates the normalized magnitude for all group velocity vectors sharing the same histogram bucket as $v_t$. Consequently, $d_i = C\left( \nabla M_i / \alpha \right)$ corresponds to a weighted polar histogram representing the distribution of group velocity for mode $M_i$. For instance, \textbf{\Cref{fig:modes}(a)-(c)} shows the polar plots resulting after applying the $C$ operator modes $M_1$, $M_2$ and $M_3$ for a unit cell of an homogeneous and isotropic material. 

In this approach the 'error'  $e$ is defined as:

\begin{equation}
e = \frac{1}{i}\sum^i_{\substack{j=2\\i>2}}
\sum^{n_\theta}_{k=1} \left( \frac{d_j^k}{d_{j-1}^k} - 1 \right)\frac{1}{n_{\theta}} 
\label{eq:err}
\end{equation}

where $d_j^k$ is the $k$-th value of $d_j(\theta)$, and $n_{\theta}$ is the number of bins used to calculate the polar histogram charts. This error metric is an average measure of the general change in directional behaviour produced by the introduction of a new term $d_i(\theta)$ into \cref{eq:dir}. The process of adding modes is stopped when the error parameter e reaches the prescribed tolerance i.e. the marginal contribution from the last mode is smaller a specified threshold.

As an example of this way to compute directional behavior, we apply \cref{eq:dir} to the unit cell of the  homogeneous and isotropic  material mentioned above. Since energy propagation in a isotropic and homogeneous material propagation occurs equally in all directions for any given frequency,  we expect to see a circle in any directionality plot.  \Cref{fig:modes}(d) shows the evolution of the polar plot describing directionality for the homogeneous material when considering $7, 14, 40,$ and $114$ propagation modes.  As the number of modes considered increases, the circular shape in the polar plot becomes more evident. In this particular case a tolerance of $0.05$ is reached after including $114$ modes. For the rest of results reported in this paper $tol$ was set to 0.05 and $n_{\theta} = 32$. The algorithm for this directional analysis approach, is fully detailed in \cref{app:dir}.

\section{Dispersive behaviour of the PXCM}
In this section, we explore the wave propagation behavior of the PXCM material from \cref{fig:PXCM1} through a series of Bloch and dynamic analyses. We restrict our attention to planar waves that lie completely in the plane of the unit cell.

We start by examining the dispersion response of a simple primitive unit cell with homogeneous and isotropic material properties. Then we modify this primitive in a sequence of steps at the end of which we recover the unit cell of the PXCM.  At each step along the way, we examine the dispersion response of the unit cell and track the evolution of the dispersion response as various structural elements of the cell are introduced.  This allows us to understand how various elements of the PXCM unit cell influence its wave propagation behavior.  Then, we compute the dispersion response of the PXCM unit cell in all three of its stable configurations. This allows us to identify partial bandgaps in its response for waves propagating along the vertical direction. We use direct integration transient analyses of a representative volume element to verify the presence of bandgaps, and to compare the response of the PXCM to harmonic excitations in the stop and pass bands. The insights gained from the foregoing analyses are then used to modify the design of the unit cell with the goal of modulating its wave propagation behavior.

The  base material forming the walls of the PXCM in this study is a generic polymer with linear isotropic properties: $E = 1$e9Pa, $\nu = 0.3$, and $\rho = 1$g/cm$^3$.  A plane strain finite element formulation is chosen for this work since we are restricting attention to planar waves. No element of the PXCM unit cell is expected to be in contact with any other element during its normal operation. Accordingly, no contact interactions between any unit cell elements are modeled in the simulations.

\subsection{Deconstruction of the wave propagation behavior of the PXCM unit cell}

\begin{figure} [H]
\subfloat{\includegraphics[width=\textwidth]{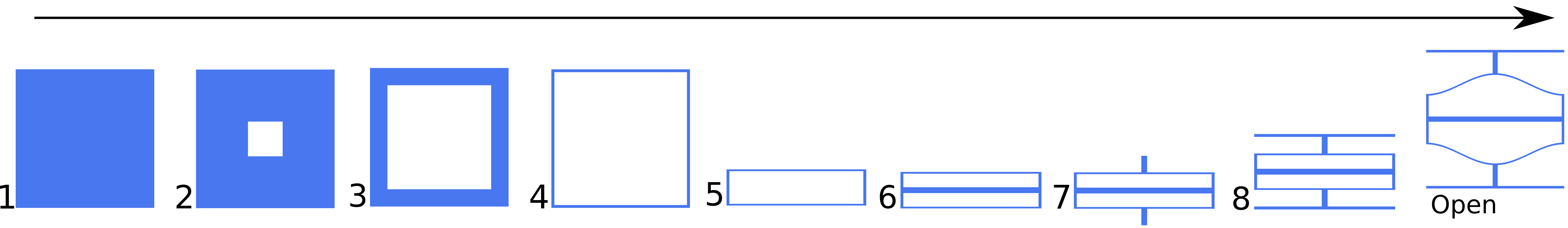}}\\
\begin{tabular}{m{0.15\textwidth}m{0.3\textwidth}|m{0.15\textwidth}m{0.3\textwidth}}
\hline
\subfloat{\includegraphics[width=0.13\textwidth]{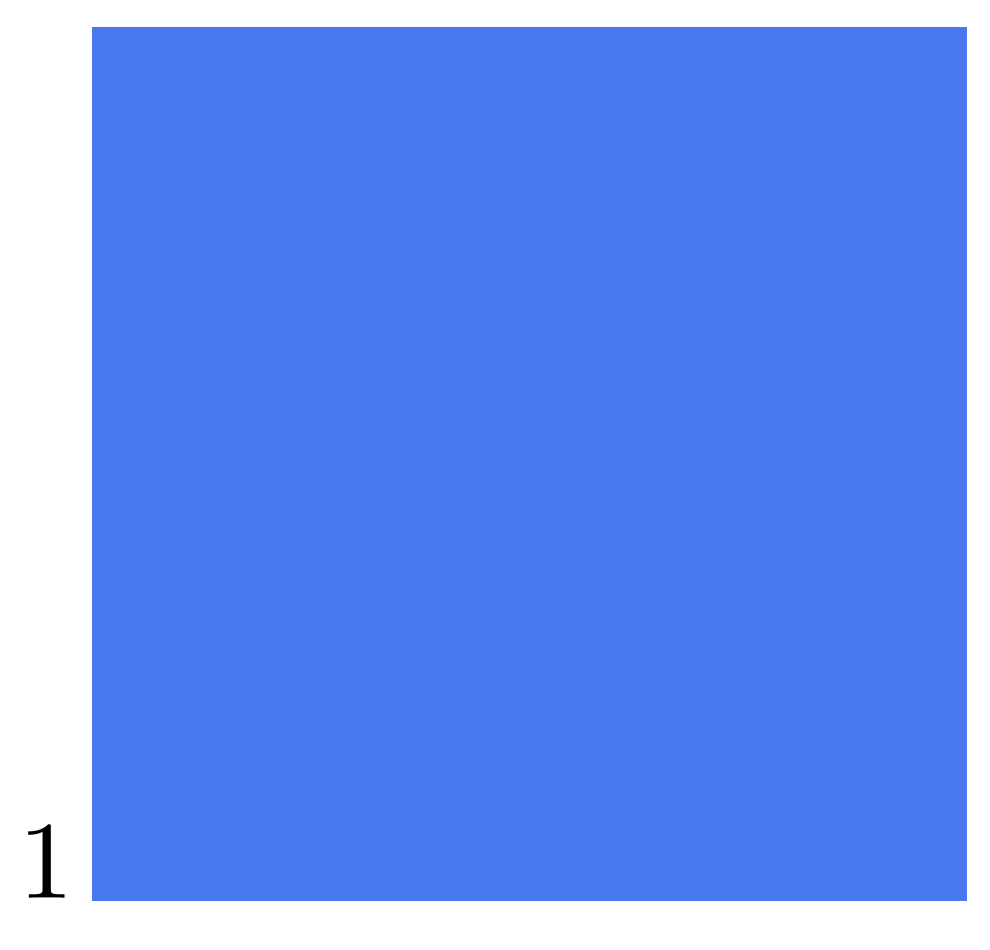}} & 
\multirow{2}{*}{\subfloat{\includegraphics[width=0.3\textwidth]{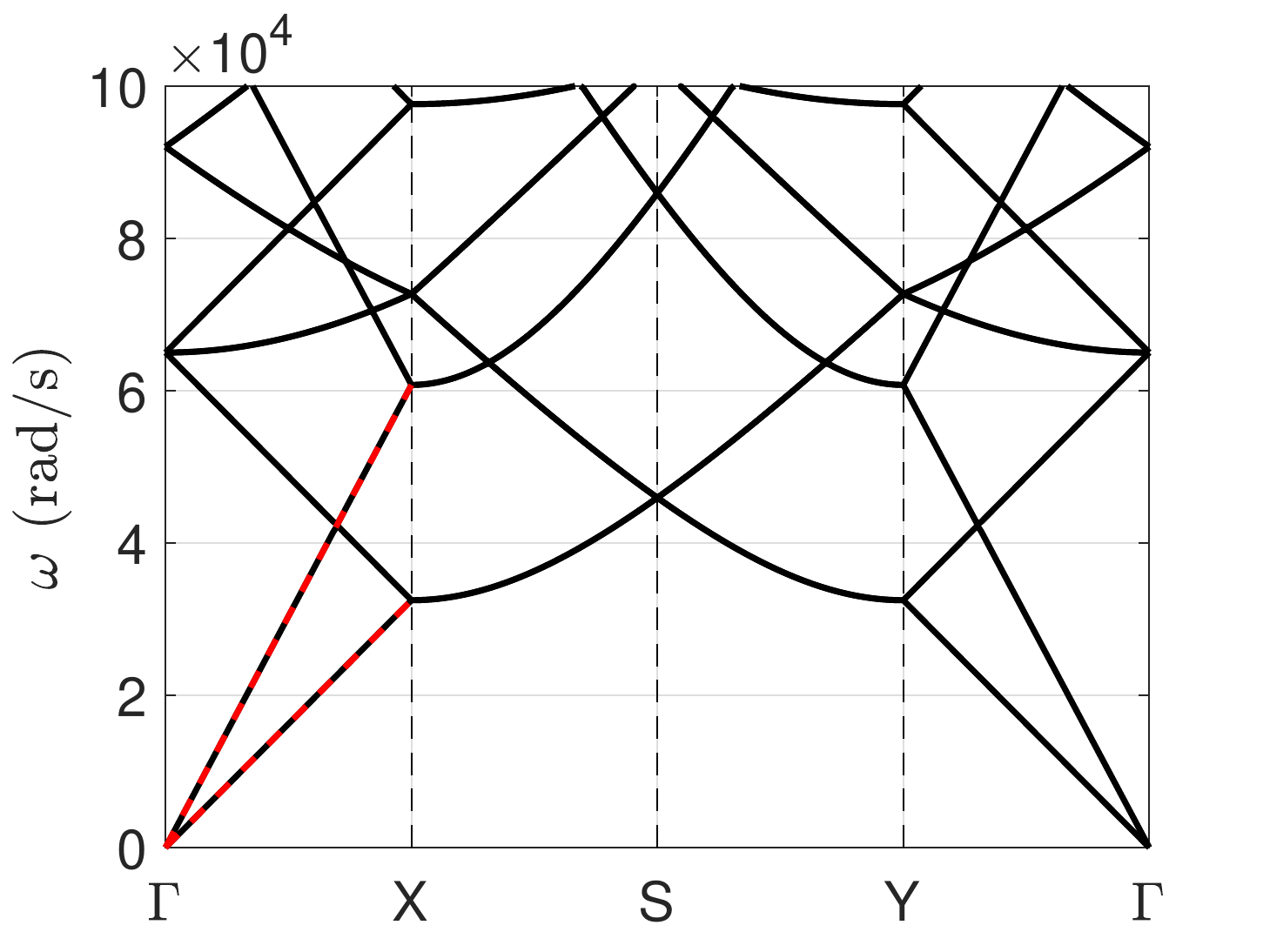}}} &
\subfloat{\includegraphics[width=0.13\textwidth]{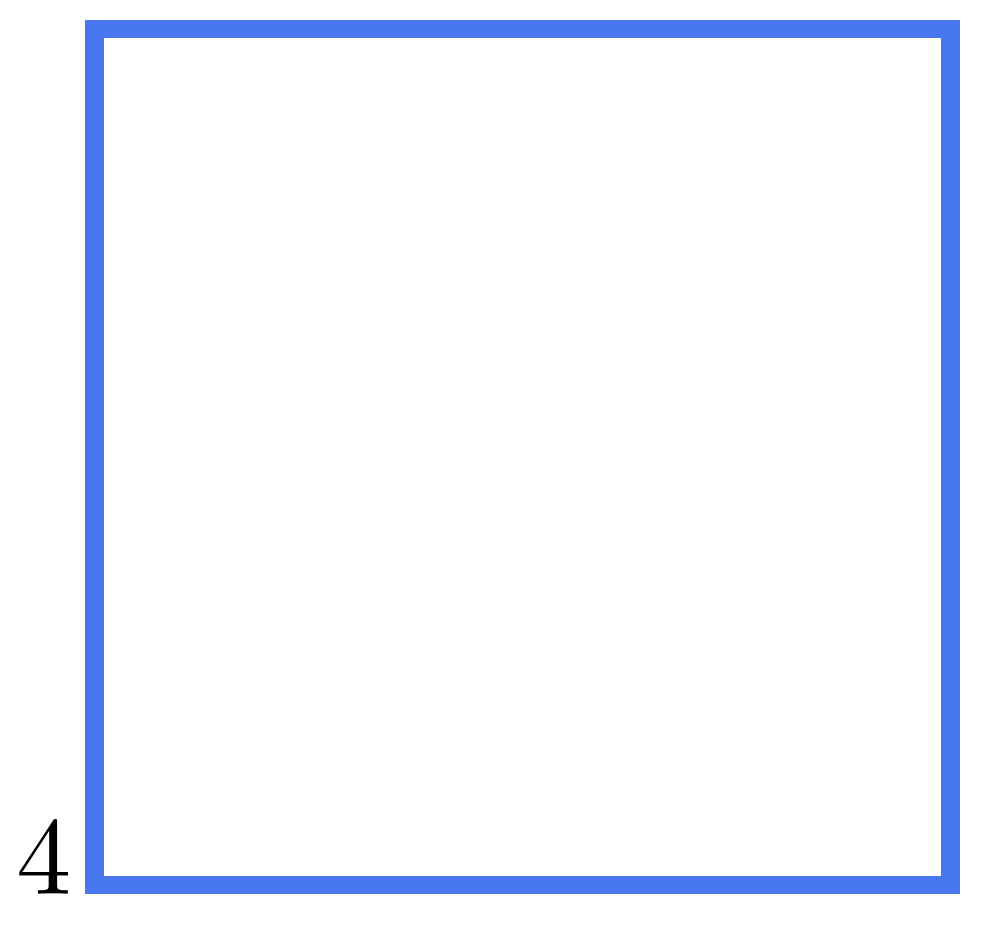}} & 
\multirow{2}{*}{\subfloat{\includegraphics[width=0.3\textwidth]{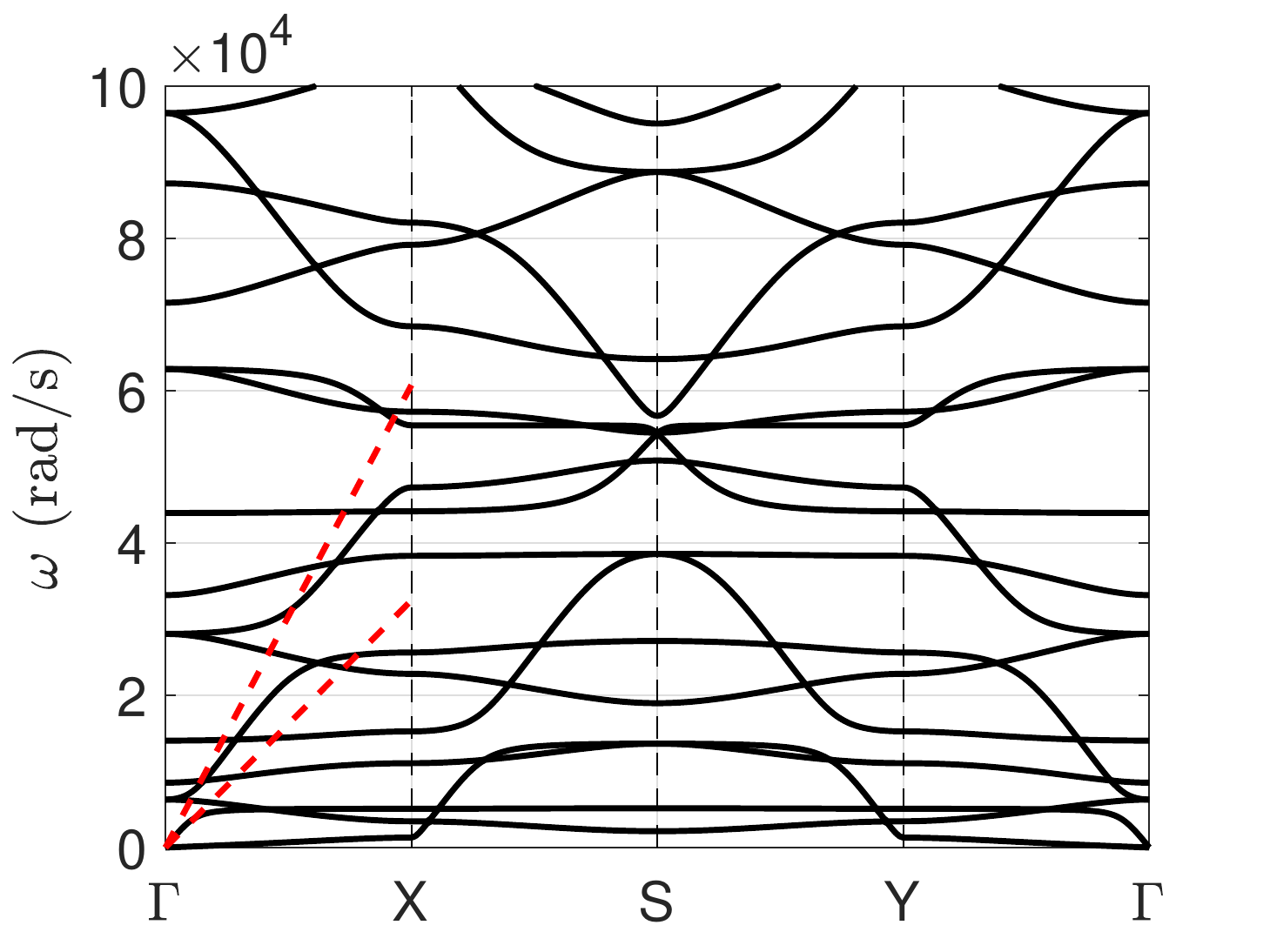}}} \\
\subfloat{\includegraphics[width=0.15\textwidth]{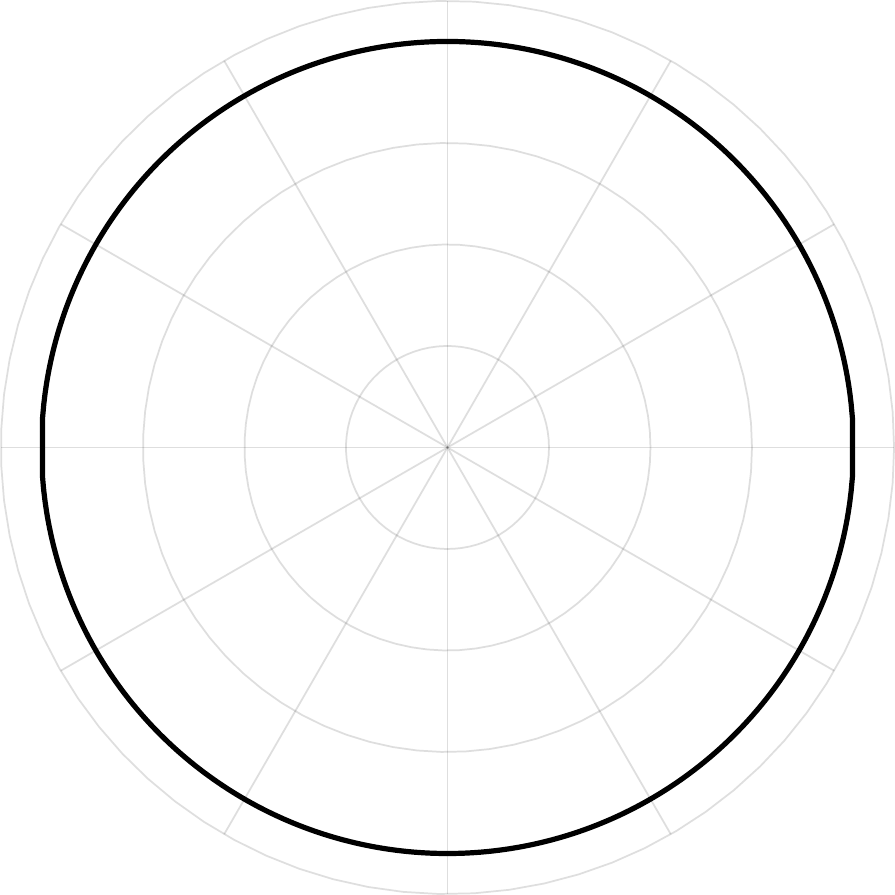}} & &
\subfloat{\includegraphics[width=0.15\textwidth]{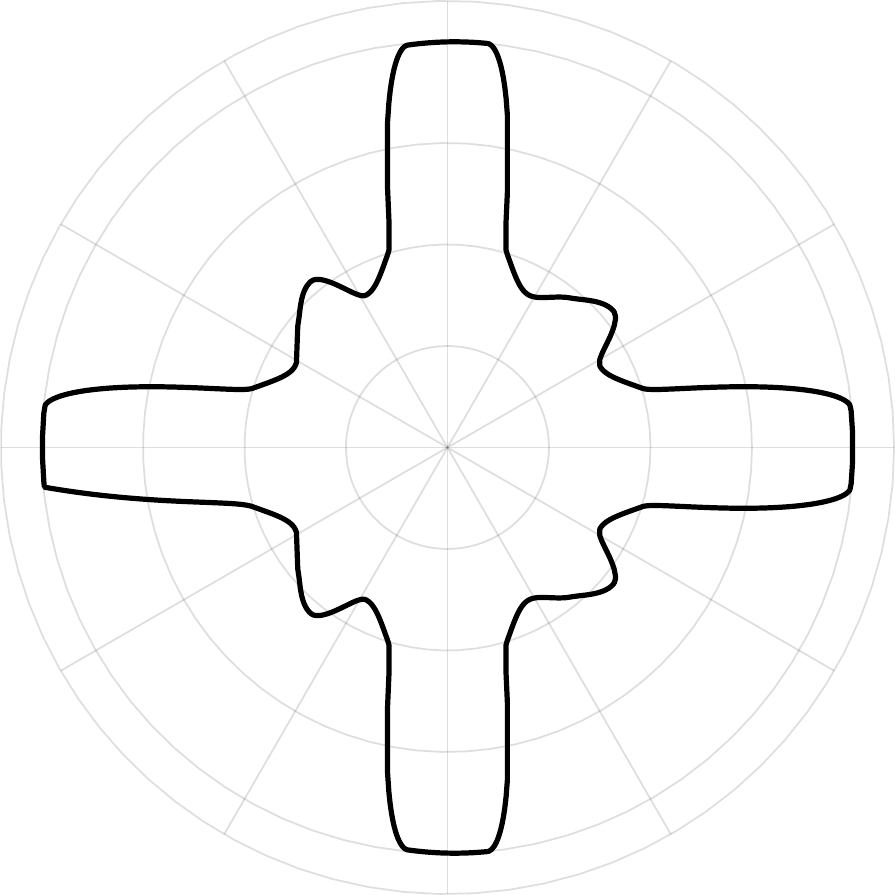}} & \\

\hline

\subfloat{\includegraphics[width=0.13\textwidth]{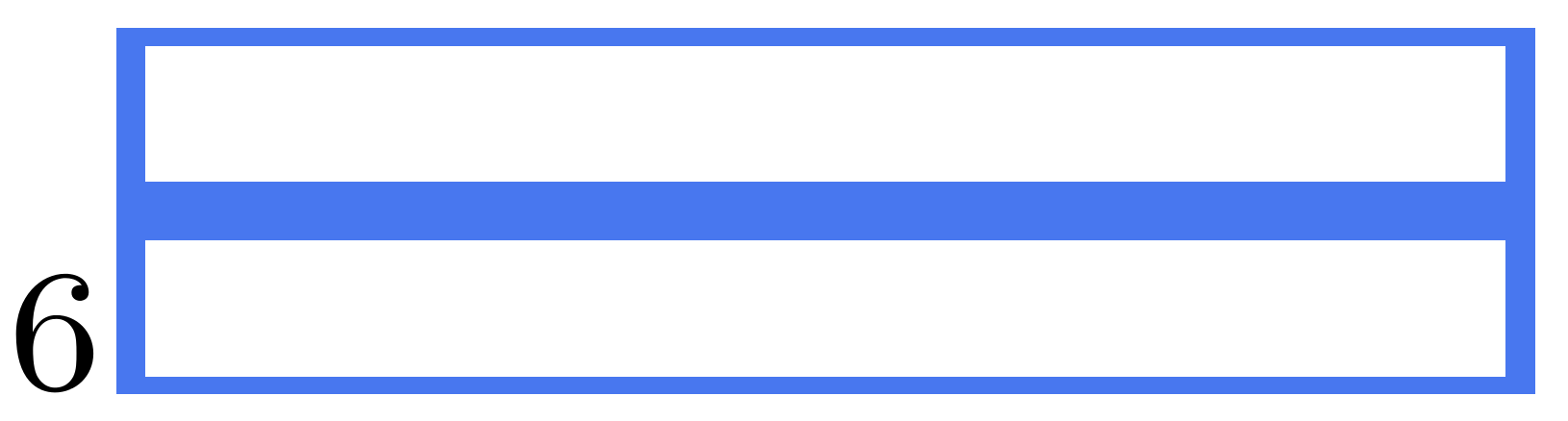}} & 
\multirow{2}{*}{\subfloat{\includegraphics[width=0.3\textwidth]{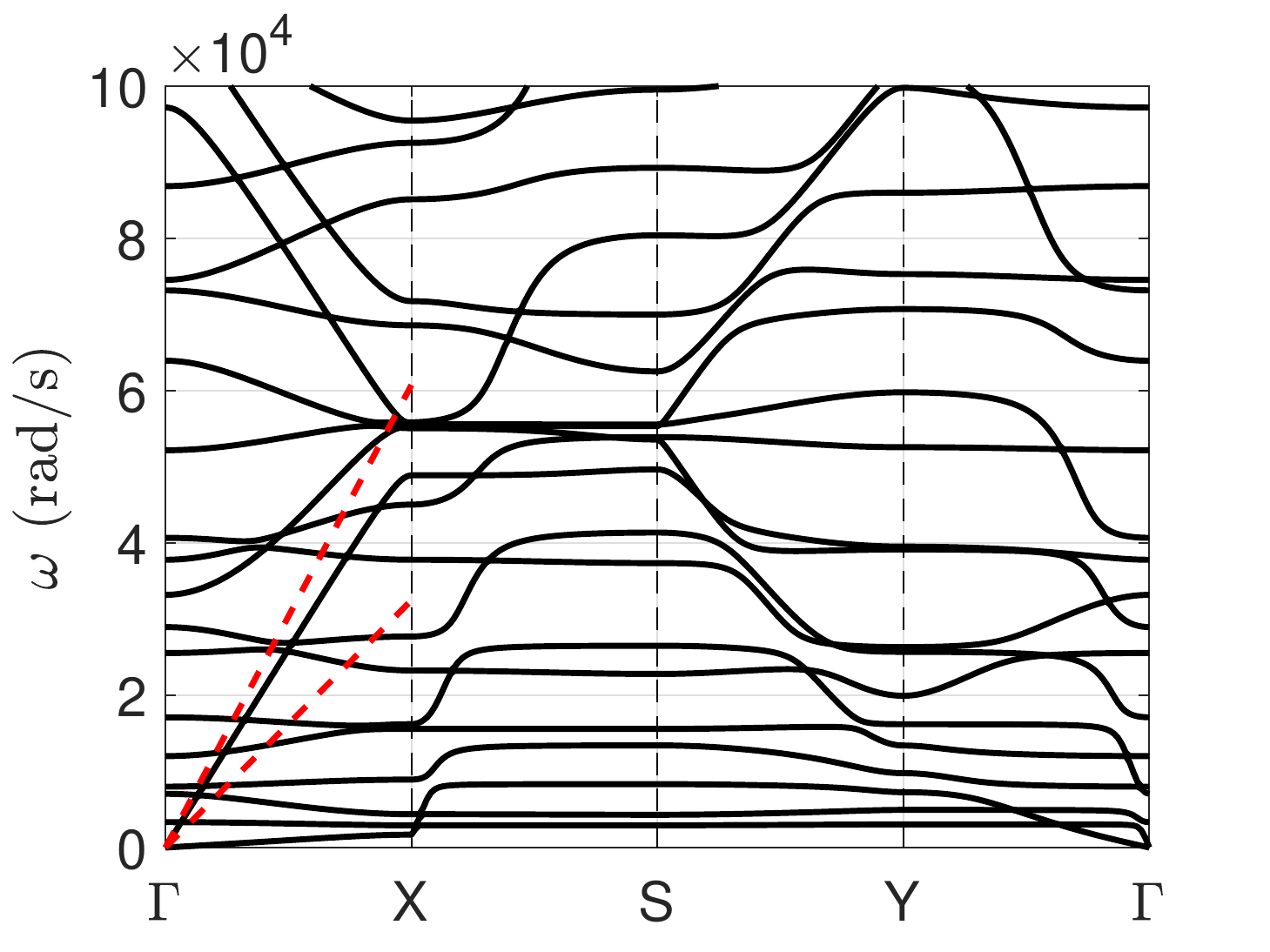}}} &
\subfloat{\includegraphics[width=0.13\textwidth]{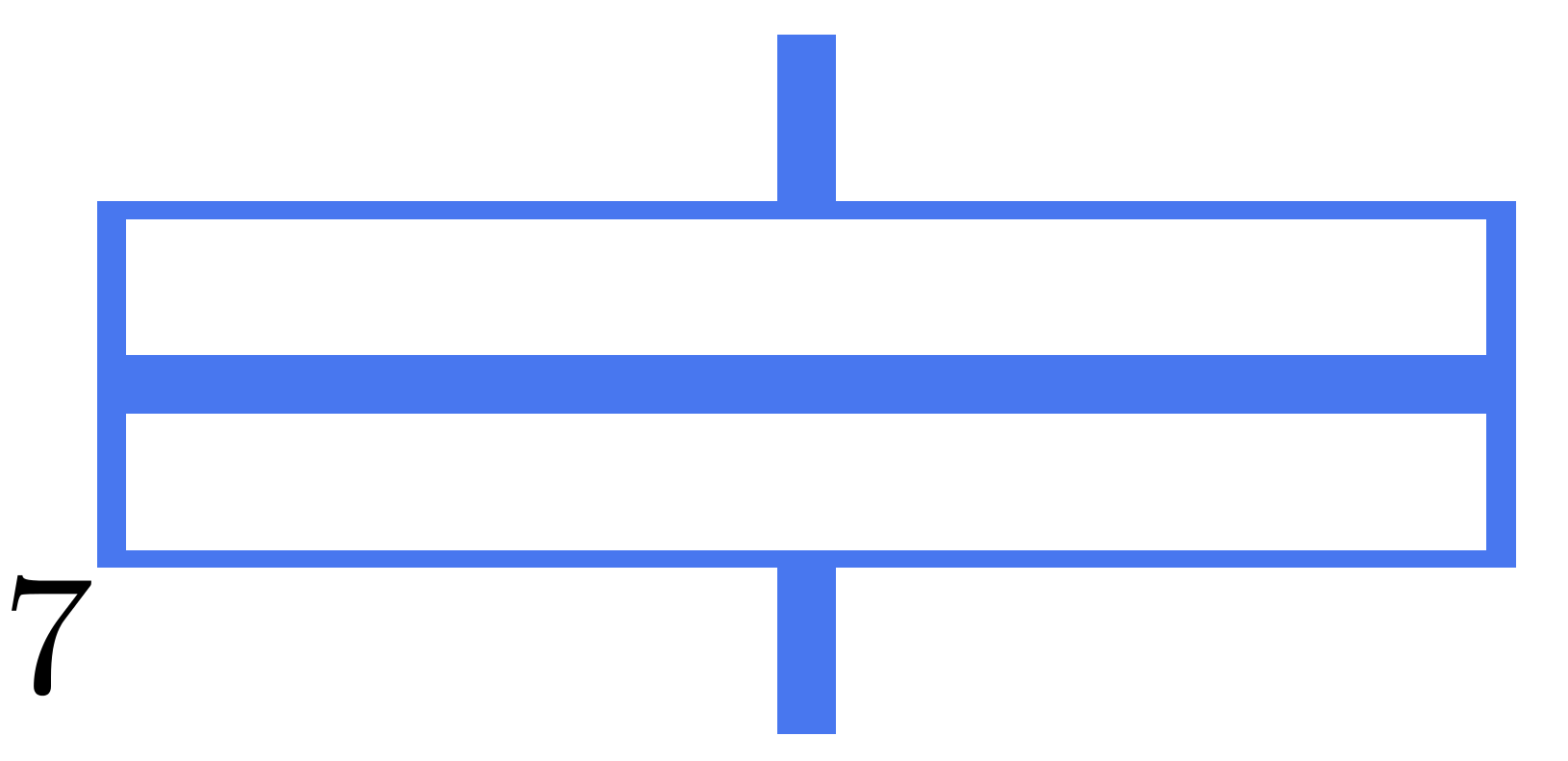}} & 
\multirow{2}{*}{\subfloat{\includegraphics[width=0.3\textwidth]{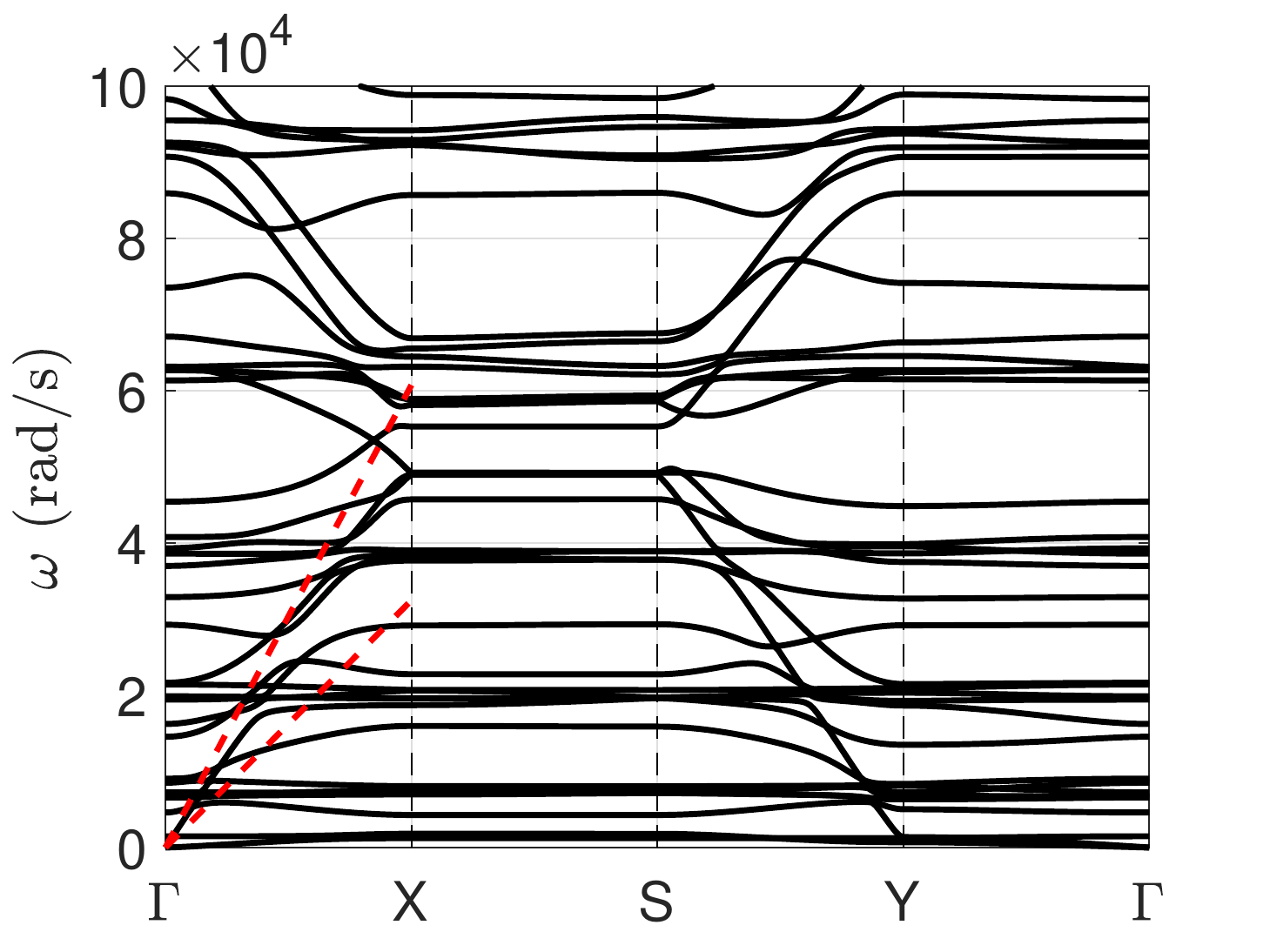}}} \\
\subfloat{\includegraphics[width=0.15\textwidth]{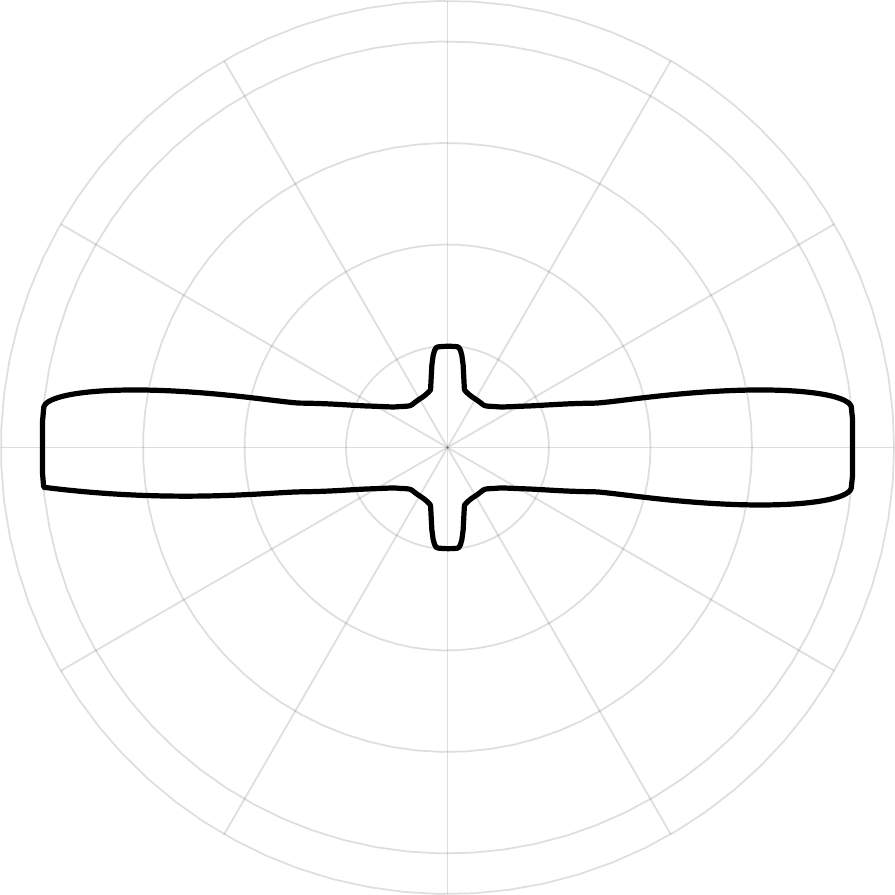}} & &
\subfloat{\includegraphics[width=0.15\textwidth]{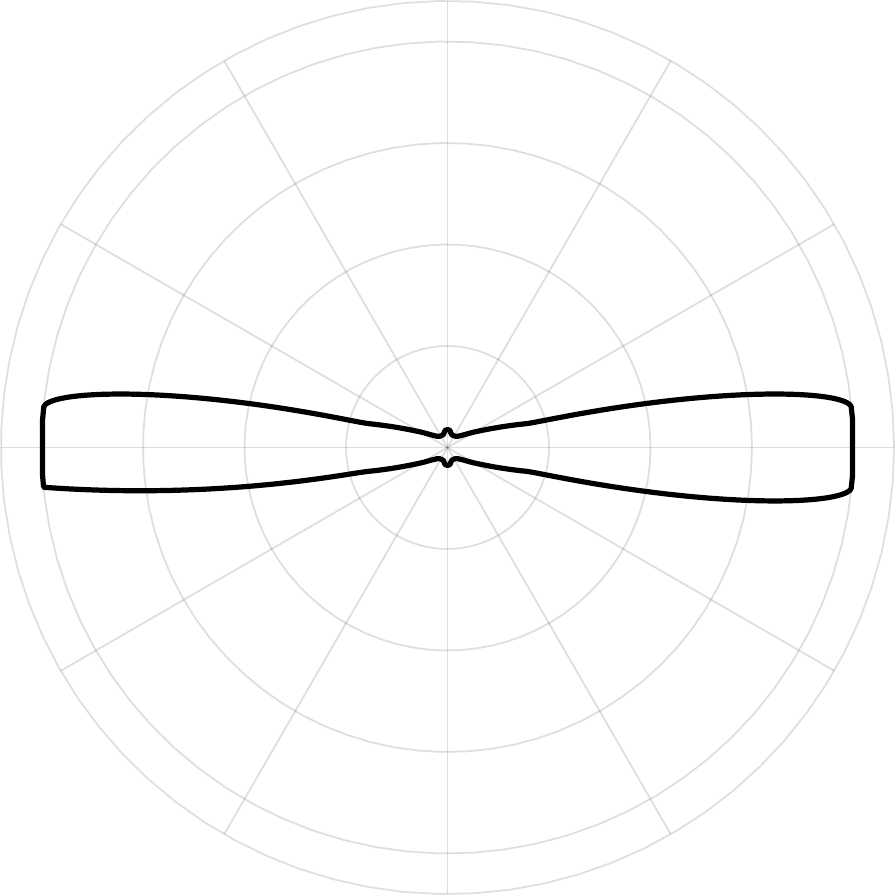}} & \\

\hline

\subfloat{\includegraphics[width=0.13\textwidth]{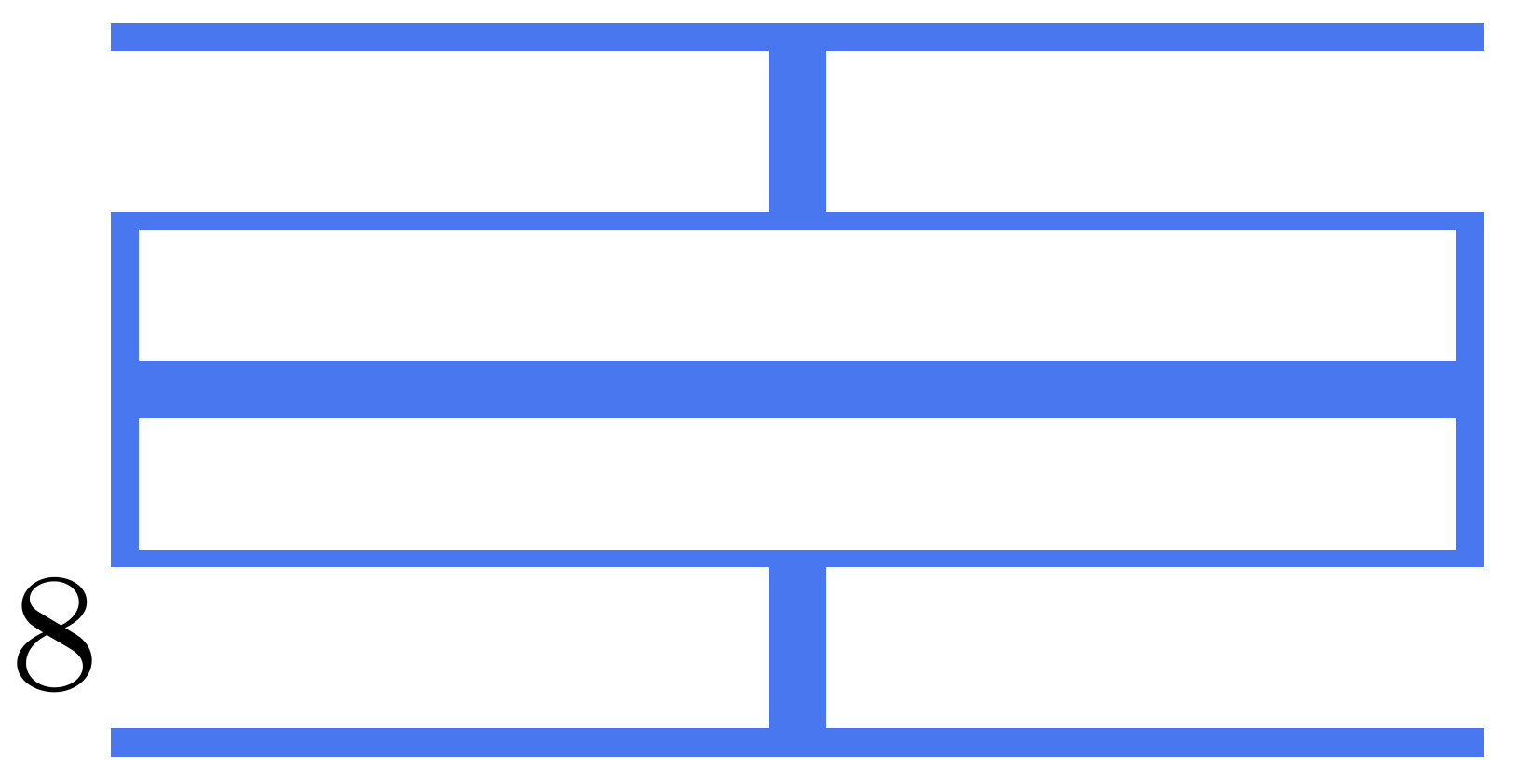}} & 
\multirow{2}{*}{\subfloat{\includegraphics[width=0.3\textwidth]{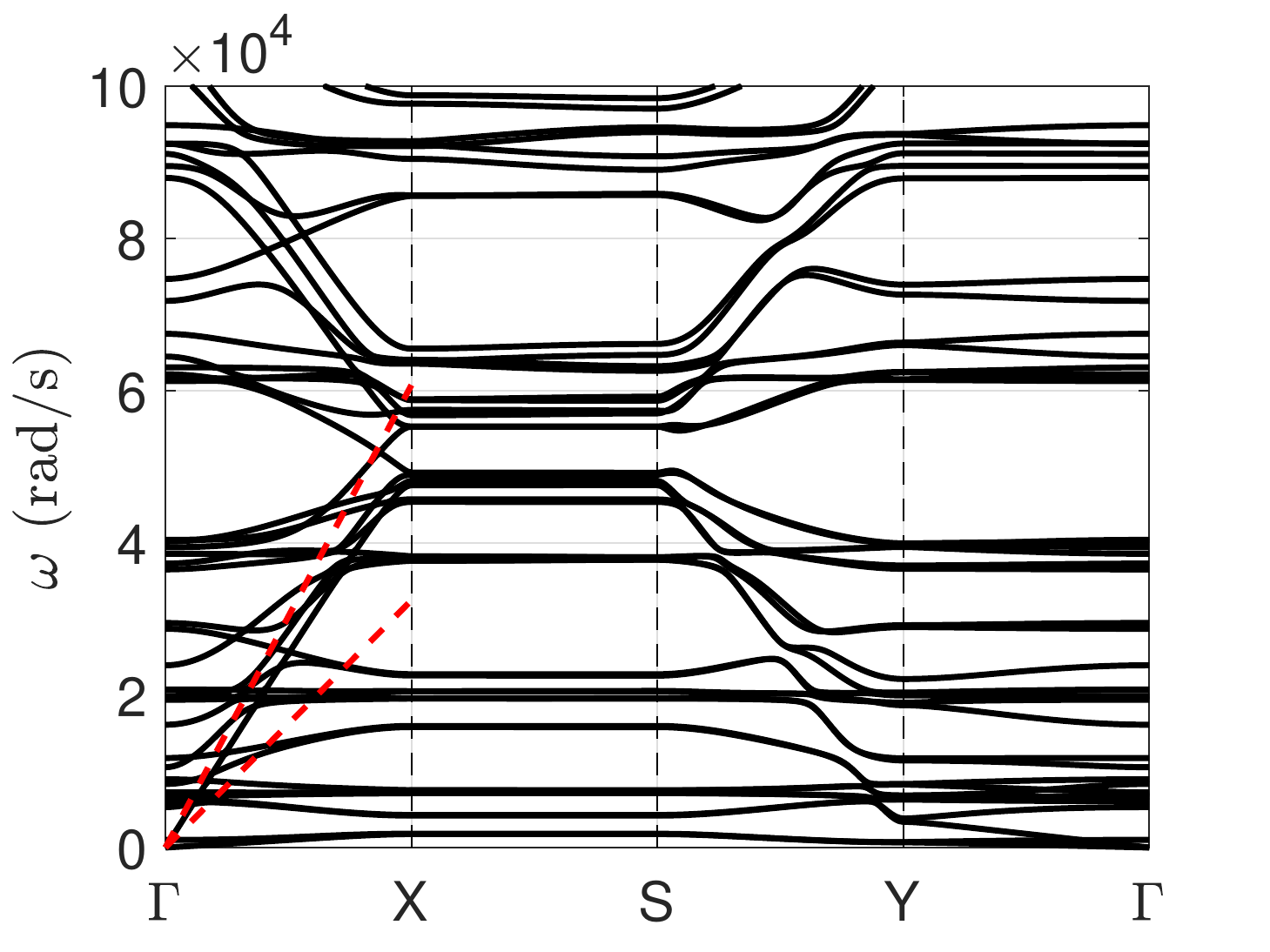}}} &
\subfloat{\includegraphics[width=0.13\textwidth]{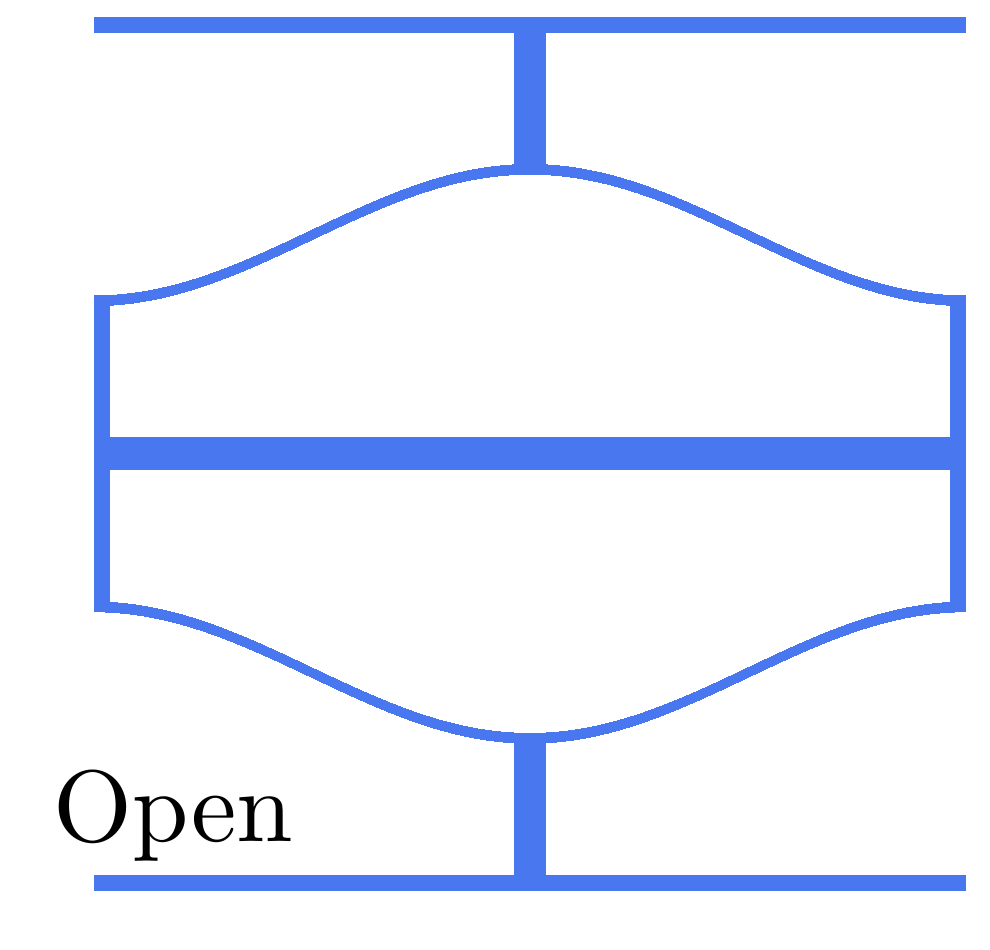}} & 
\multirow{2}{*}{\subfloat{\includegraphics[width=0.3\textwidth]{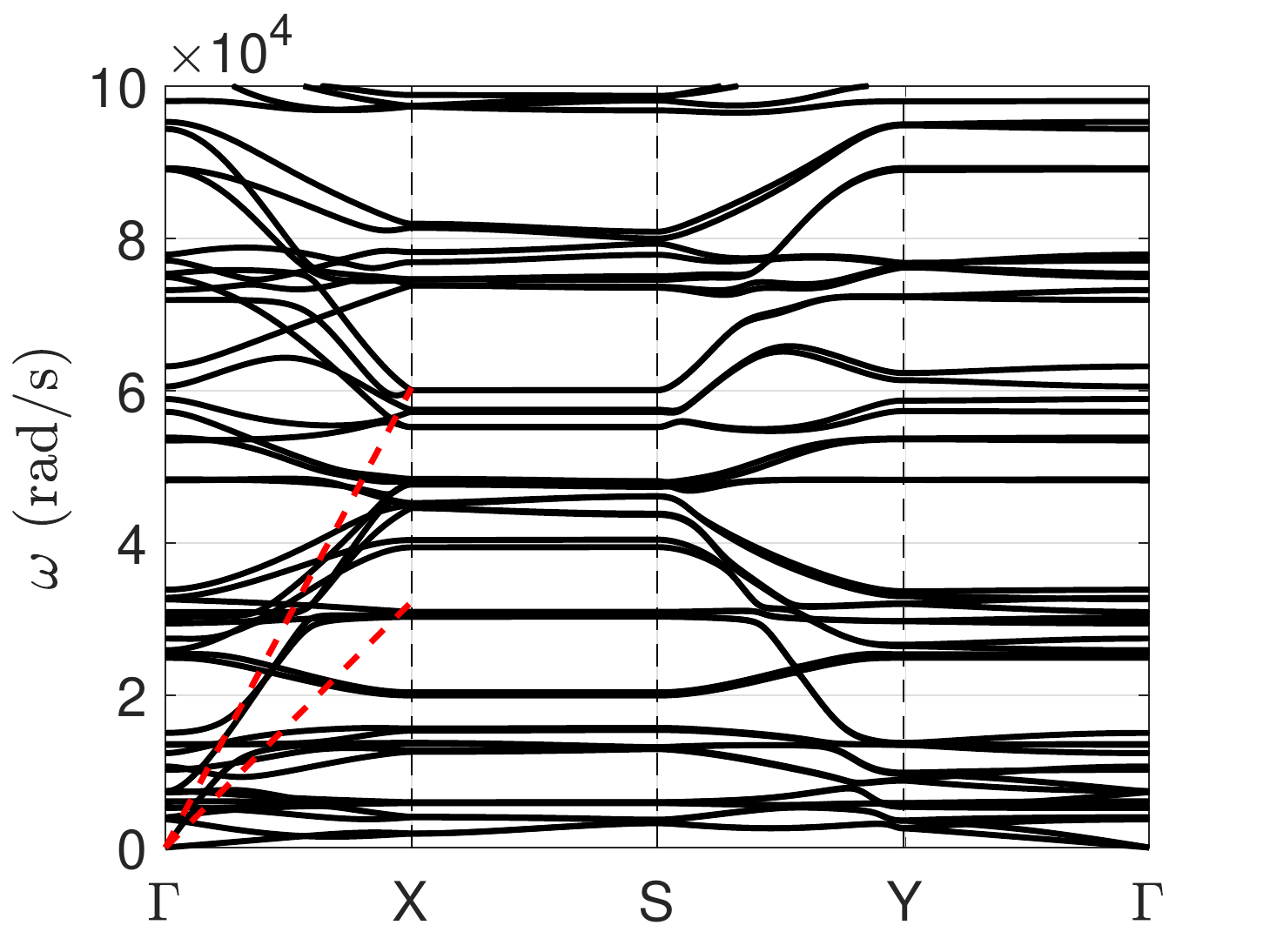}}} \\
\subfloat{\includegraphics[width=0.15\textwidth]{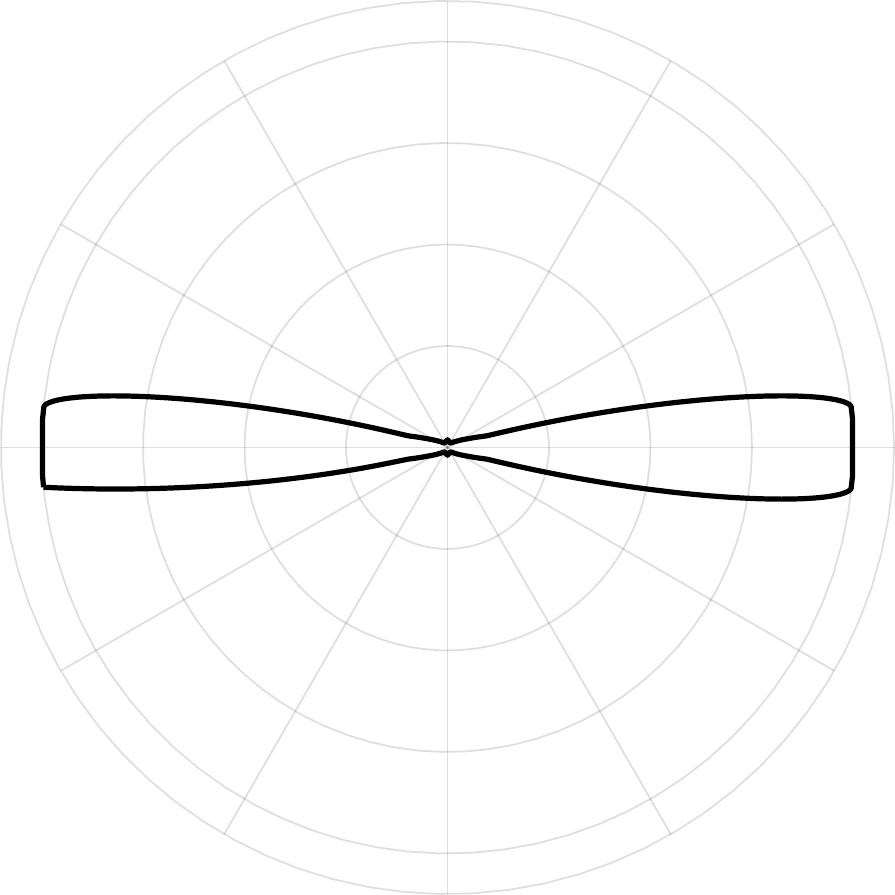}} & &
\subfloat{\includegraphics[width=0.15\textwidth]{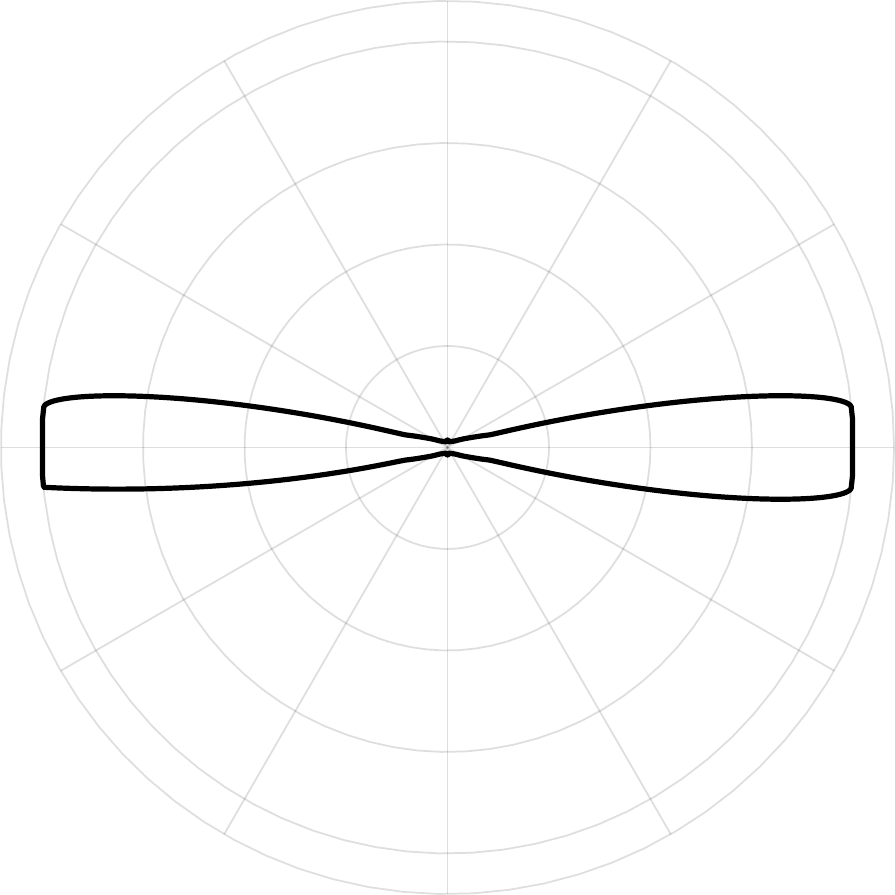}} & \\
\hline
\end{tabular}
\caption{Progressive construction of the unit cell. In each frame: unit cell geometry, dispersion diagram and directionality of energy propagation. In the dispersion diagrams, slopes of the dashed red lines overlaid for reference in the dispersion diagrams indicate the P- and S-wave speeds in the homogeneous material. For briefness purposes, only some salient stages of the decostruction are depicted.\label{fig:prog}}
\end{figure}

\textbf{\Cref{fig:prog}} shows the evolution of the geometry beginning with a homogeneous isotropic primitive and ending with the final open configuration of the PXCM.  We begin with a square unit cell made of a homogeneous isotropic material in Stage 1.  The geometric simplicity of this configuration also gives rise to a simple wave propagation behavior, which makes this cell a useful reference. The geometry of the unit cell, its dispersion diagram and its directionality plot are shown in \cref{fig:prog}. The stage 1 unit cell can only sustain $P$ (longitudinal) and $S$ (shear) planar waves and, as expected, these waves propagate uniformly in all directions in this cell as seen in the directionality plot. No complete or partial bandgaps are visible in the dispersion diagram. Dashed red lines are overlaid on dispersion diagram. The slopes of these lines yield the magnitudes of group velocities for the P and S waves respectively. The magnitude of the group velocity is a measure of the speed of energy propagation in the material via these waves.  These lines will be overlaid on the dispersion diagrams for the cells at successive stages to facilitate a comparison of how quickly energy propagates in the various cell designs. Note that $\Gamma X$ direction includes all the symmetries present in the homogeneous material unit cell. In order to track the variations in wave propagation properties during this construction process, the contour of the IBZ from the PXCM unit cell (i.e. $\Gamma X S Y \Gamma$) is used. 

A small square pore is introduced in the center of the unit cell for Stage 2.  The size of this pore is gradually increased through Stage 4, where the unit cell is reduced to a thin frame. A few observations are noteworthy in the evolution from stage 1 to stage 4.  The complexity of the dispersion diagram increases significantly during the transition. Now the unit cell is capable of sustaining not just longitudinal and shear planar waves but also bending planar waves. It is evident from the dispersion diagrams along the $\Gamma X$ and $Y \Gamma$ directions that the magnitude of group velocities for the P and S waves decreases as the size of the pore increases. This is a consequence of the wave packages encountering less material to propagate.  Furthermore, along the $\Gamma X$ direction this decrease occurs faster for the S-wave than for the P-wave. The magnitude of the group velocity of the P-wave for the stage 4 cell is still close to the reference value from the stage 1 cell, while the magnitude of the group velocity for the S-wave is reduced to $5\%$ of its reference value from stage 1. This is explained by presence of horizontal bars along the $\Gamma X$ direction that have a large longitudinal stiffness in comparison with their transverse stiffness. An analogous response is observed along the $Y \Gamma$ direction. The change in material distribution in the unit cell gives rise to a dramatic change in the energy propagation behavior of the unit cell in stage 4.  The uniform propagation of energy in all directions that was observed in the stage 1 cell, is replaced by a highly anisotropic directional behavior that favors energy propagation along horizontal and vertical directions.

At Stage 5,  we change the aspect ratio of the cell so that the square cell of stages 1--4 is now morphed into a rectangular one whose width is much greater than its height. This destroys the symmetry in the dispersion diagram about $S$, which was a recurring feature in stages 1 -- 4. A stiff horizontal bar is introduced at the center of the unit cell  at Stage 6. The introduction of yet another structural member in the unit cell gives rise to a further increase in the complexity of the dispersion diagram. The newly added horizontal bar is significantly stiffer in its longitudinal direction than either the top and bottom horizontal bars or the vertical side bars. This makes the horizontal direction the preferred direction for energy propagation for the unit cell at stage 6.

We define a material as being continuous in a specific direction if a ray can follow a single uninterrupted path when travelling along that direction. If such a path exists, the material is continuous in that direction and a wave would be able to propagate in the direction independently of the value of $\kappa$. Unit cells in stages 1 -- 6 exhibit material continuity in the horizontal and vertical directions,  therefore there are neither complete (i.e. cover the entire IBZ) nor partial band gaps in the dispersion diagrams for these cells. Vertical bars are added at the centers of the top and bottom horizontal bars in Stage 7. Topologically, the main difference between the unit cell at stage 7 and the cells in the previous stages is that the vertical bars added in stage 7 disrupt the material continuity in the vertical direction. This gives rise to partial band gaps for waves propagating along the diagonal ($X S$) and vertical ($Y \Gamma$) directions. The stiffness of the central horizontal bar ensures that the stiffness of the cell in the horizontal direction continues to dominate its stiffness in the vertical direction. This is reflected in the horizontal direction continuing to be the preferred direction for energy propagation in the unit cell.

A stiff horizontal bar is added at the top and bottom of the cell in Stage 8.  This results in a further strenghtening of the preference for energy flow in the horizontal direction.  The design of the unit cell at the last stage (Stage 9) is topologically identical to the design at stage 8, with the only difference being that two of the horizontal bars that were straight in stage 8 are now bent into a sinusoidal shape in stage 9.  There are no discernible differences in either the dispersion diagrams or the directionality plots for the unit cells in stages 8 and 9.  The unit cell in stage 8 has a stable monotonic force displacement relationship for moderate forces applied to the vertical bars at the top and bottom of the cell. The change in the shape of some of the beams in the cell in going to the stage 9 cell fundamentally alters the force displacement relationship by introducing non-monotonicity that is driven by an elastic instability. This endows a material based on the unit cell in stage 9 to exhibit solid state energy dissipation.  However, despite the significant differences between these two unit cell designs in terms of their static mechanical responses, their wave propagation behavior is largely indistinguishable because their topologies are identical. In this section, only the results from some salient values of the decostruction are shown. For completeness, results from all stages are included in the supplementary material.

Based on the above discussion, we can draw some broad conclusions about the role of the various structural elements in the unit cell of the PXCM.  The unit cell has many horizontal ($\Gamma X$ direction in the wave vector space) bars with high longitudinal stiffness. These bars favor the propagation of P-waves, while they oppose that of S-waves. We observe branches in the dispersion diagram along the $\Gamma X$ direction with high group velocities corresponding to P-waves.  Thus, the stiff horizontal bars in the unit cell favor the transport of energy via P-waves along the horizontal direction and slow down the propagation of the S-mode.
The introduction of the stiff vertical bars at the centers of the top and bottom horizontal bars  in stage 7 not only disrupts the material continuity of the unit cell in the vertical direction, but it also signifcantly reduces its vertical stiffness. Therefore, beginning with stage 7 we notice  a considerable decrease in the magnitude of the P- and S-wave group velocities along the vertical  ($Y \Gamma$) direction.  The disruption in the material continuity of the unit cell gives rise to partial band gaps  along the $X S$ and $Y \Gamma$ directions in the dispersion diagram.
Finally, we note that while changes in shape of the structural elements can affect the static mechanical behavior of a unit cell, these changes do not noticeably alter the wave propagation behavior of the unit cell if its toplogy is unchanged.
While these observations are based on the analysis of a particular unit cell, the underlying concepts apply more generally to multi-stable periodic cellular materials. We will use these concepts throughout the rest of this section as we further explore the the wave propagation behavior of the PXCM studied here.

\subsection{PXCM cell in different stable configurations}
\begin{figure}[H]
\begin{tabular}{m{0.25\textwidth}m{0.35\textwidth}m{0.35\textwidth}}
\centering
\subfloat{\includegraphics[width=0.25\textwidth]{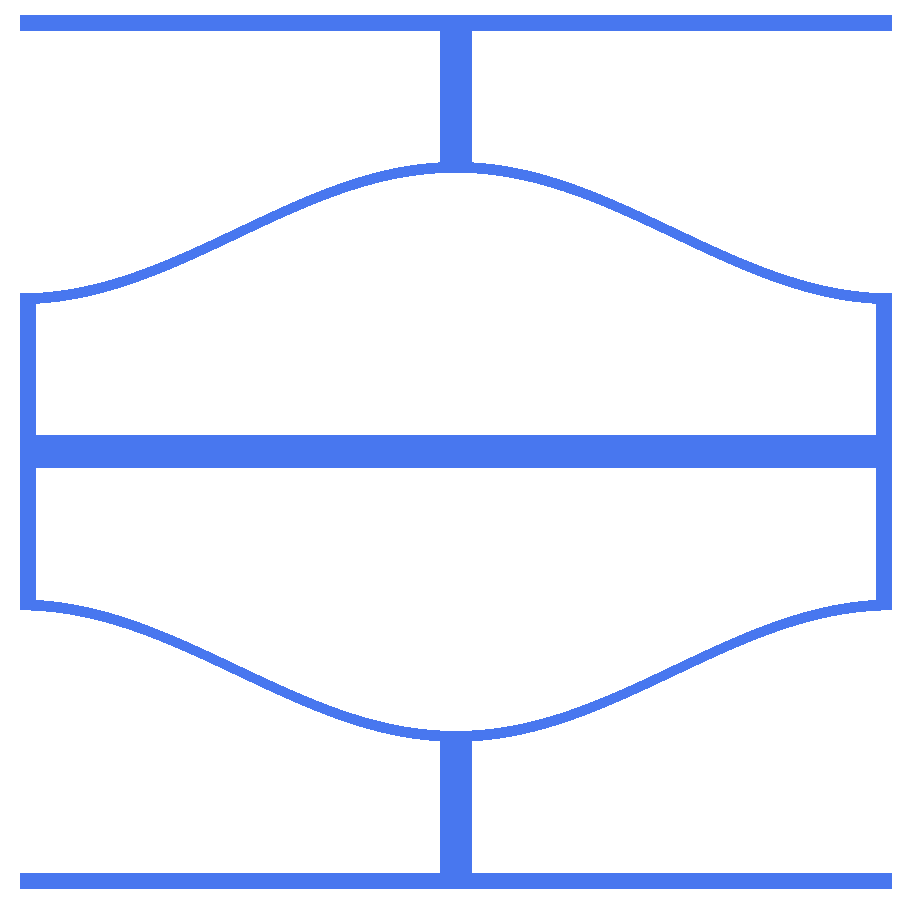}} &
\subfloat{\includegraphics[width=0.35\textwidth]{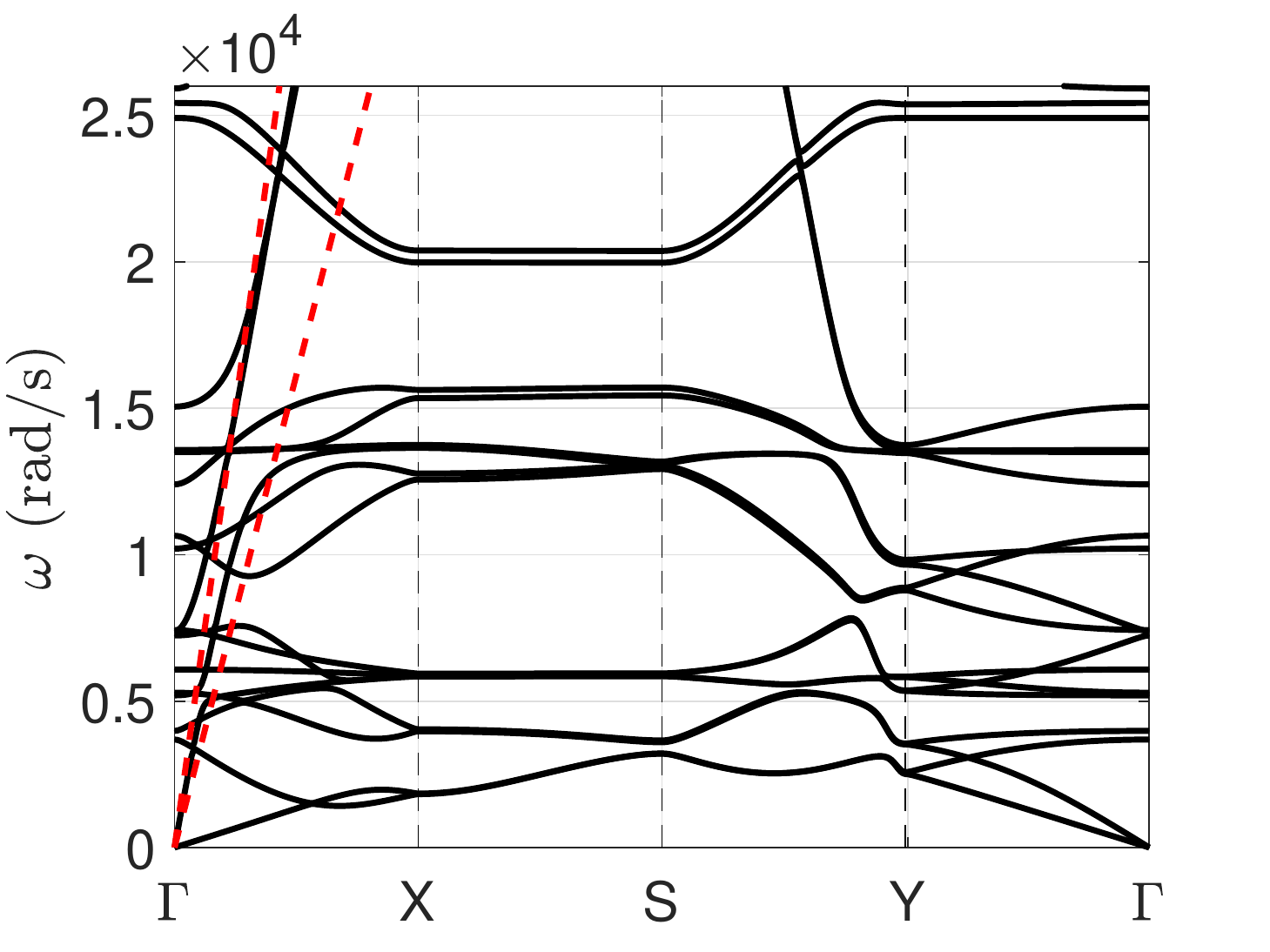}} &
\subfloat{\includegraphics[width=0.35\textwidth]{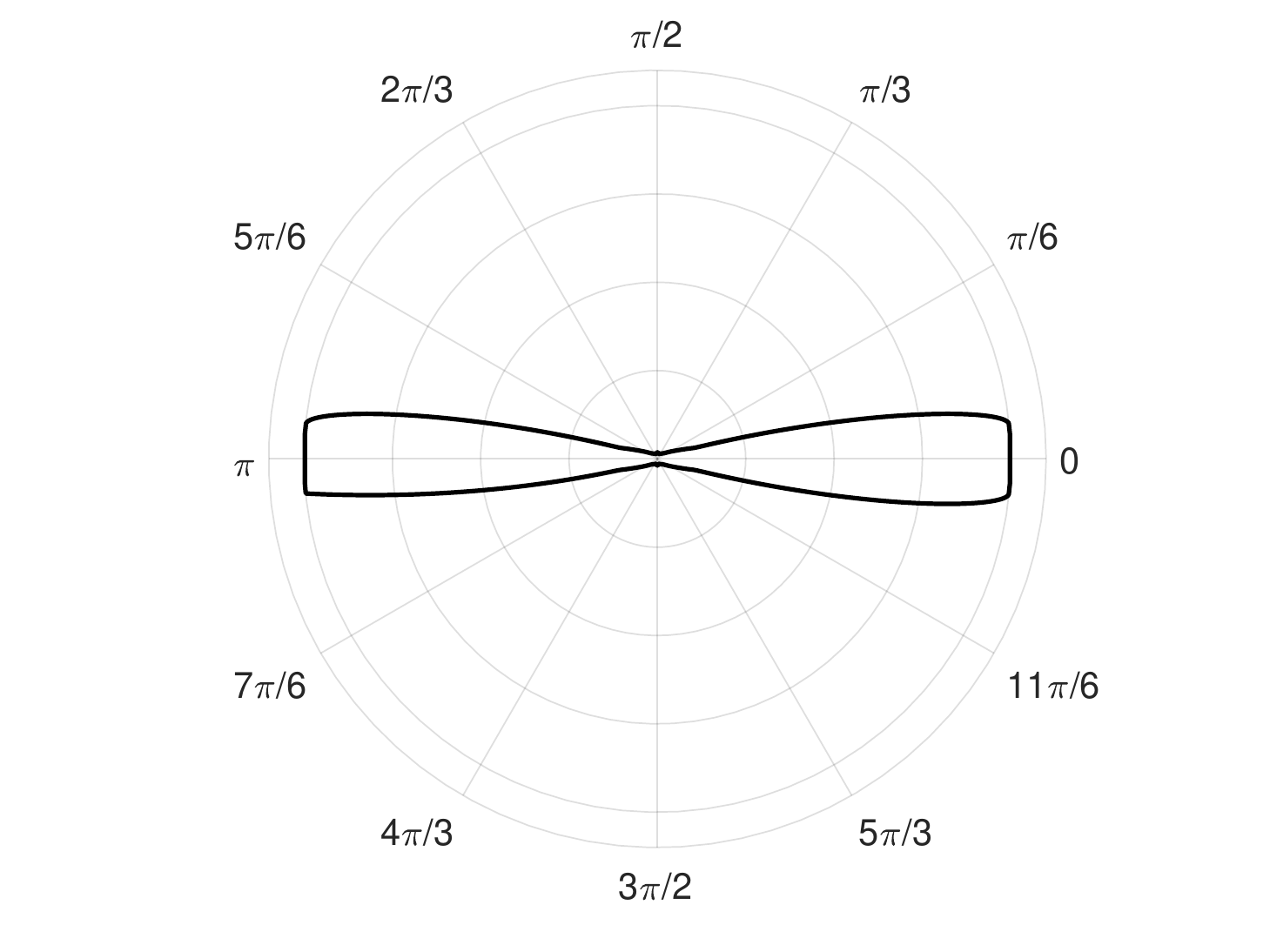}}\\
\subfloat{\includegraphics[width=0.25\textwidth]{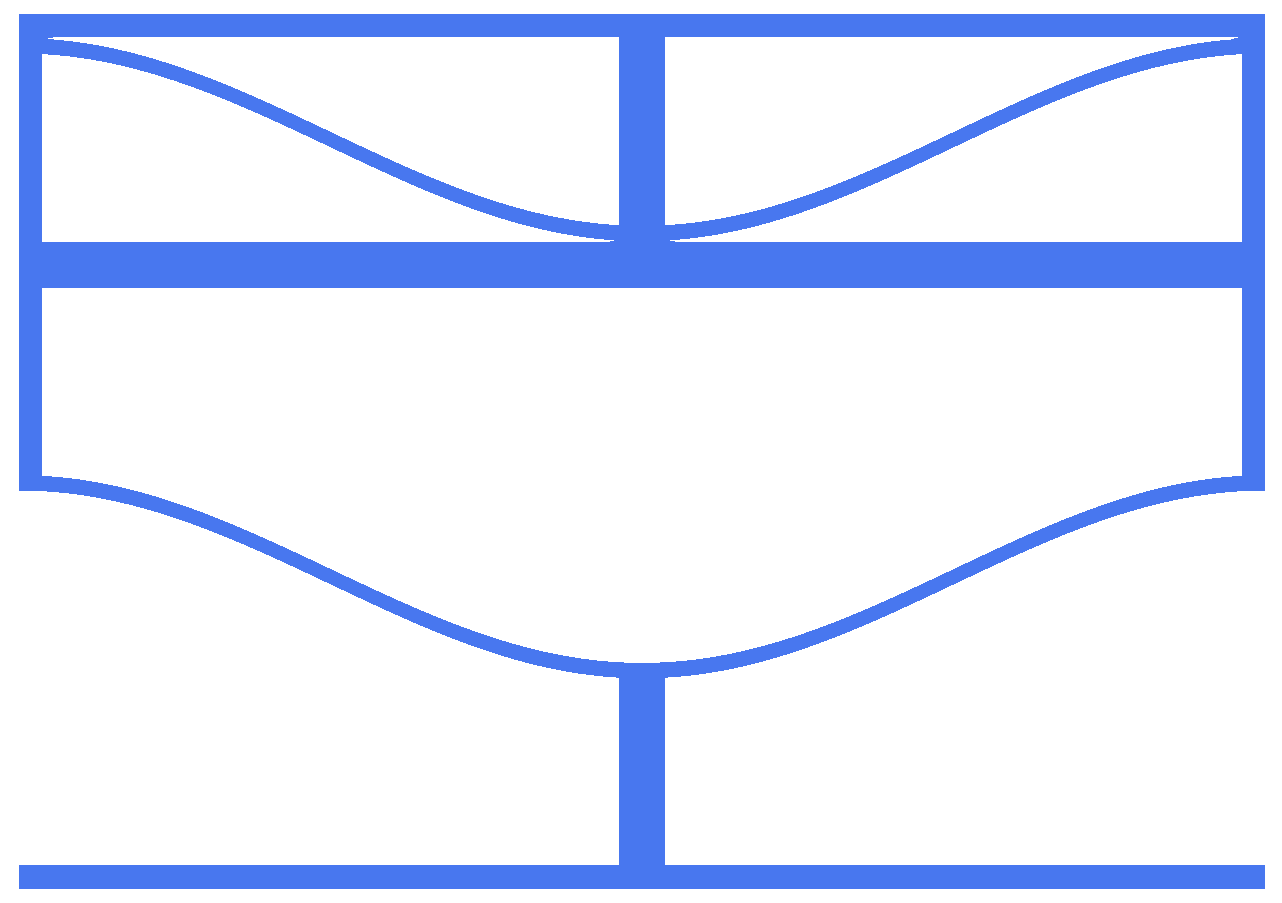}} &
\subfloat{\includegraphics[width=0.35\textwidth]{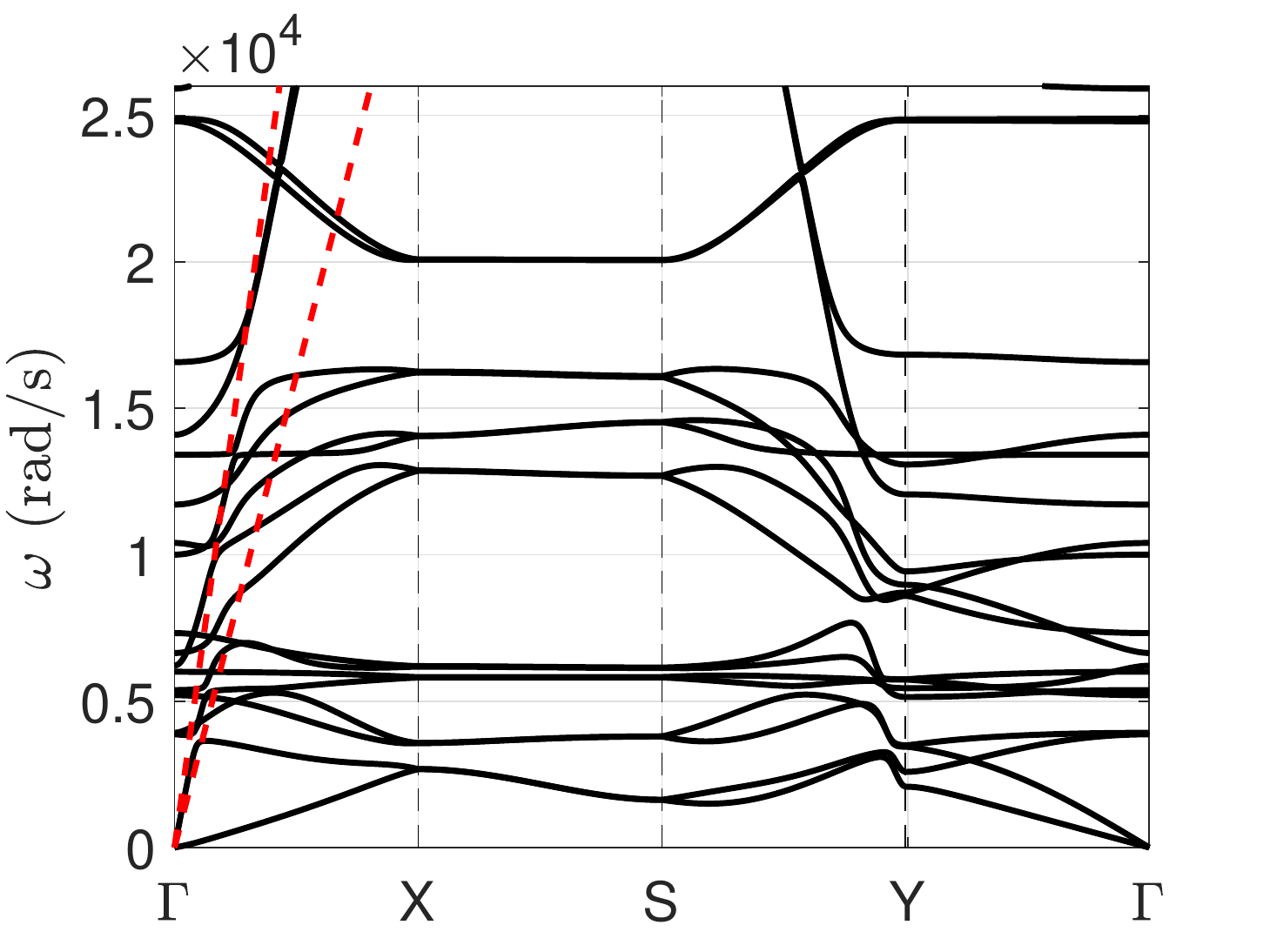}} &
\subfloat{\includegraphics[width=0.35\textwidth]{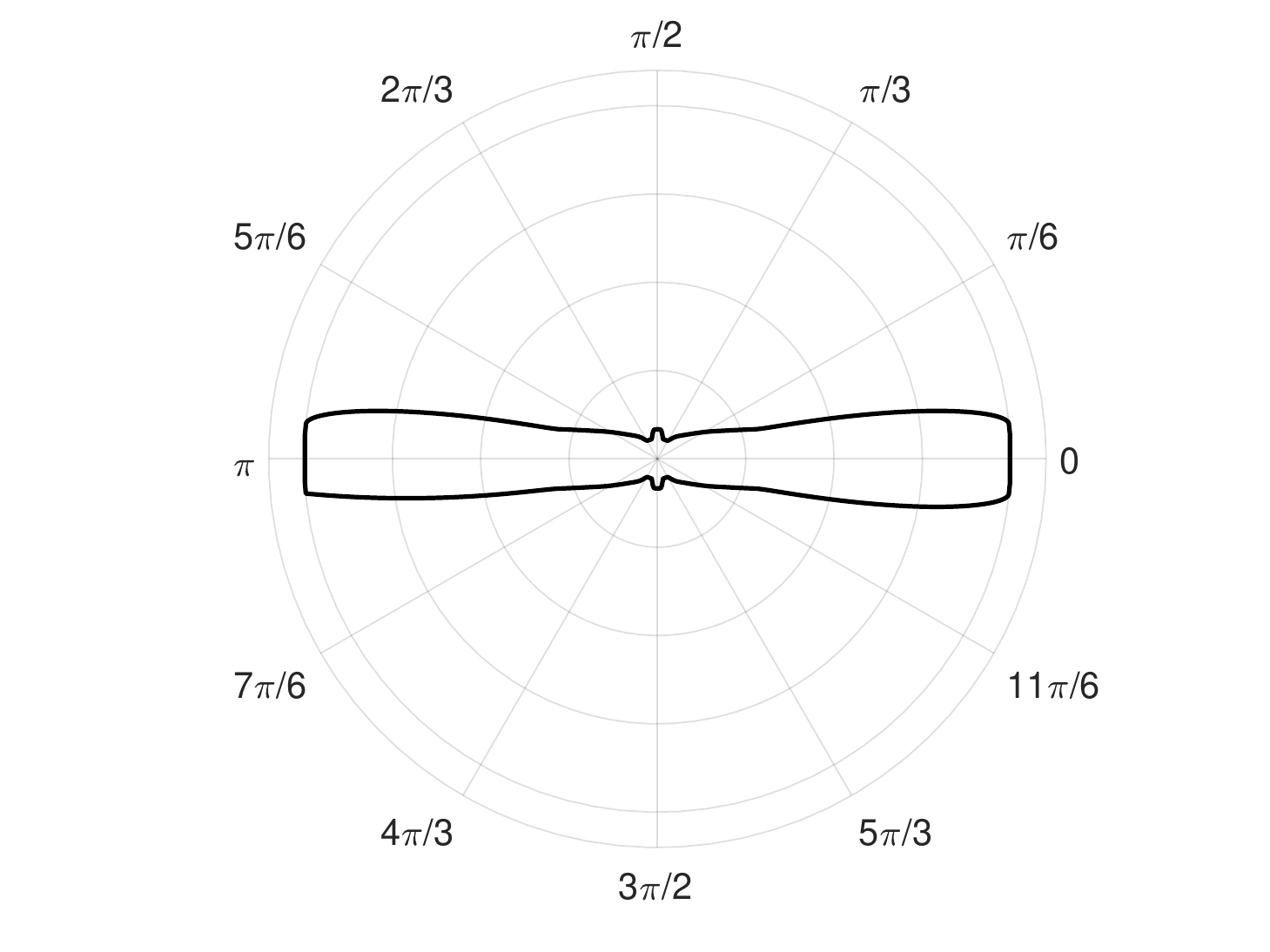}}\\
\subfloat[(a)]{\includegraphics[width=0.25\textwidth]{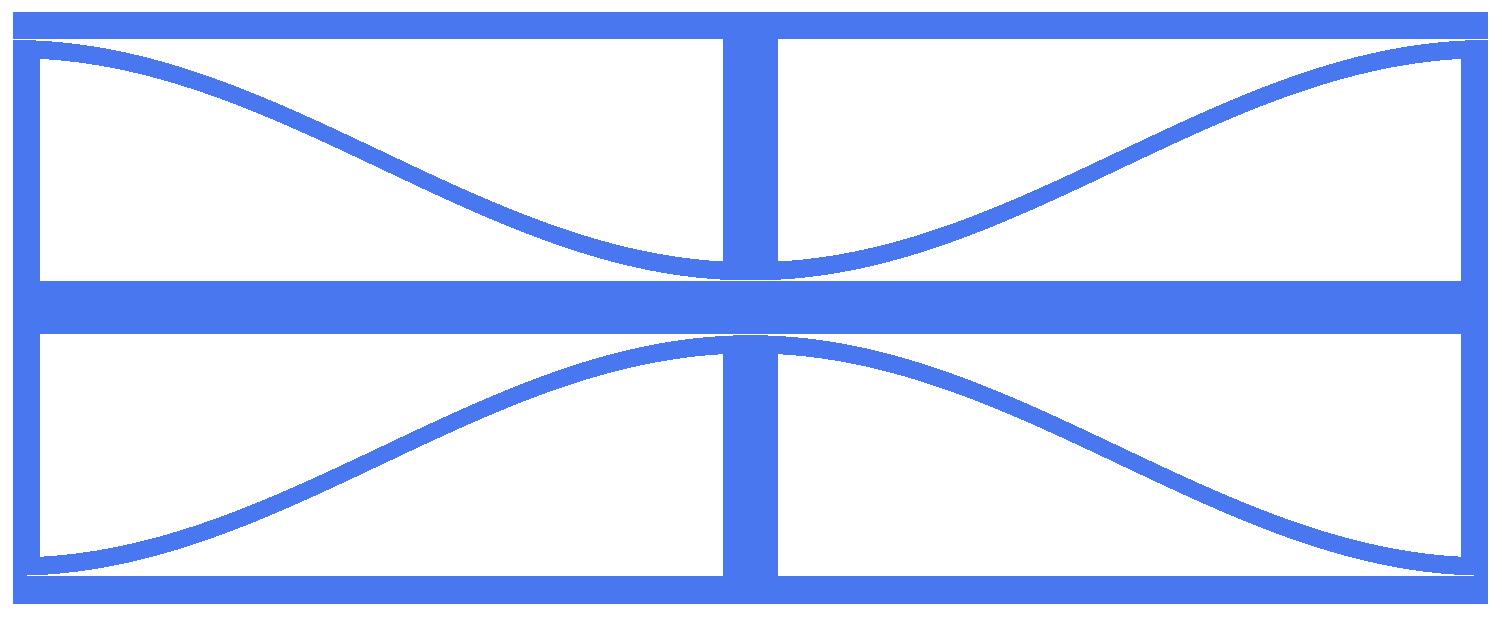}} &
\subfloat[(b)]{\includegraphics[width=0.35\textwidth]{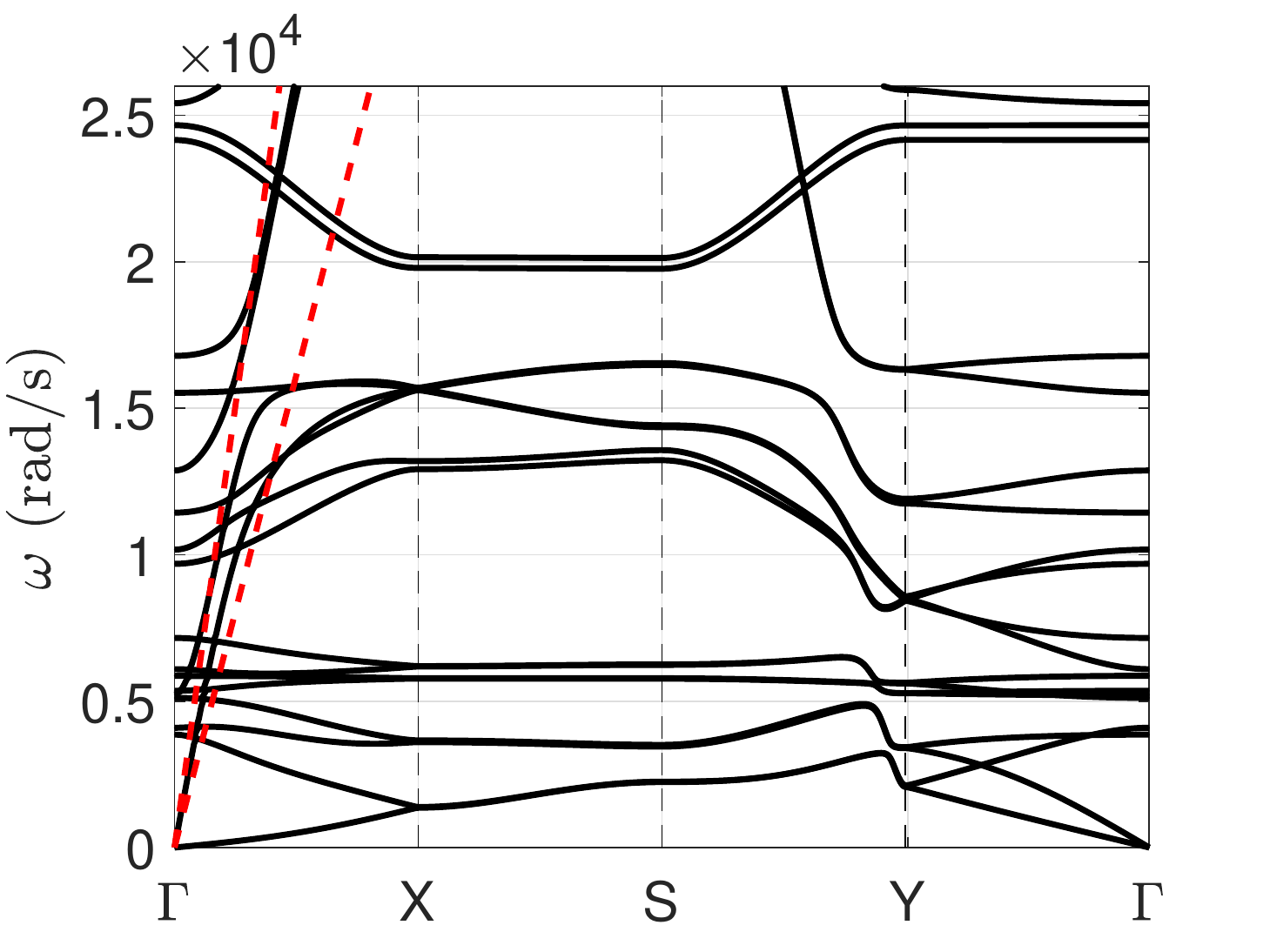}} &
\subfloat[(c)]{\includegraphics[width=0.35\textwidth]{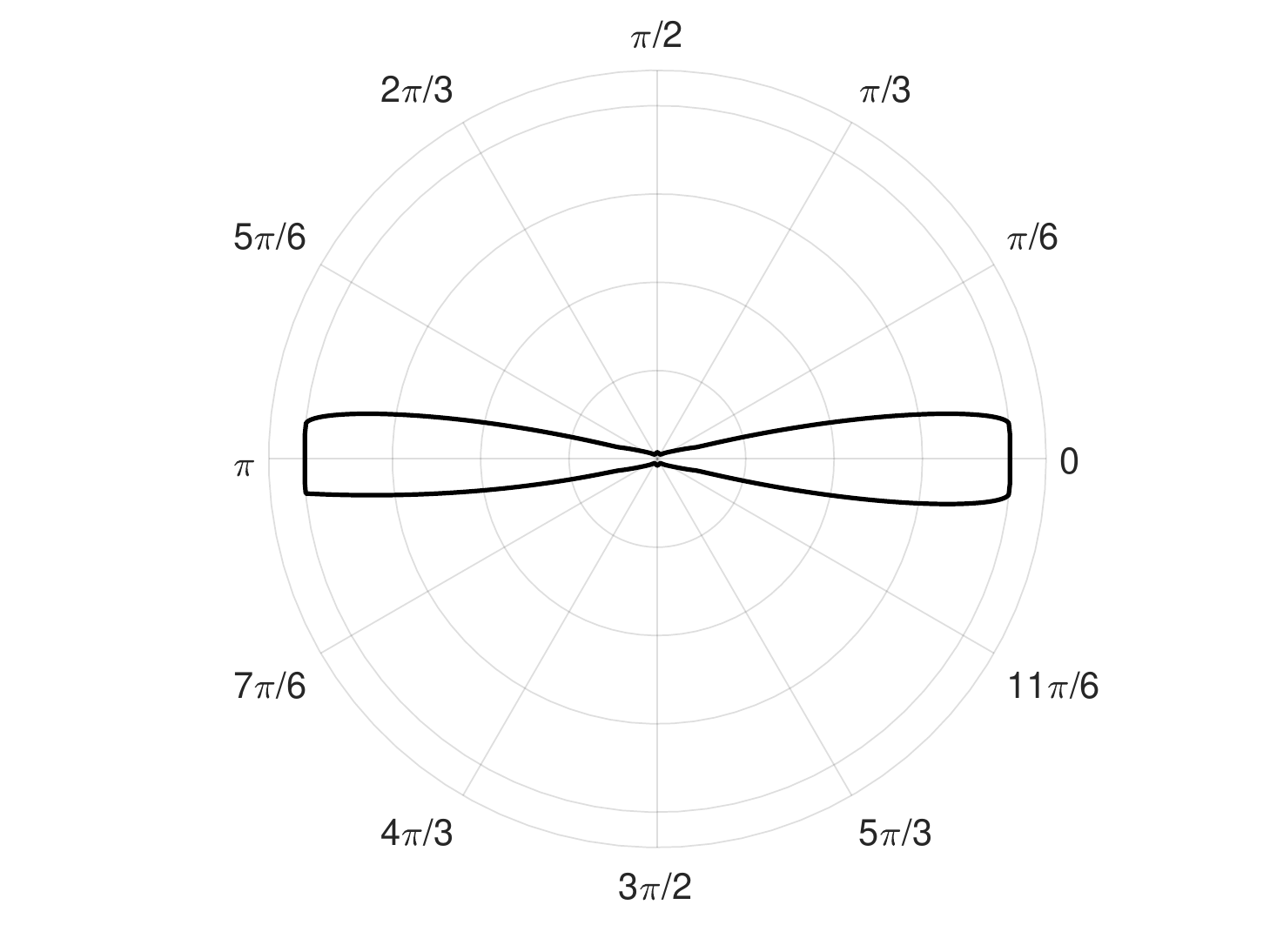}}
\end{tabular}
\caption{Wave propagation behavior of the PXCM unit cell in its three unique configurations. Column (a) - configuration geometry, Column (b) - dispersion diagram and Column (c) - directionality plot. Dashed red lines are overlaid on the dispersion diagrams to indicate the magnitudes of the P- and S-wave group velocities in the homogeneous material.}
\label{fig:comp}
\end{figure}

An unit cell of the PXCM comprises two multi-stable mechanisms. These can be both bistable, both metastable or a mixture of the two. This gives rise to three unique configurations for the unit cell: the open configuration, the closed configuration and an intermediate one. These are shown in column (a) of figure \textbf{\cref{fig:comp}}. We note that despite the significant differences in the geometries of the two configurations of the underlying mechanism the corresponding topologies are identical. Therefore, based on the learnings from the previous discussion, we anticipate that the wave propagation behavior of the three different unit cell configurations would be very similar. 

The dispersion diagrams (column b) and directionality plots (column c) for the three unit cell configurations are shown in \cref{fig:comp}.  As expected, the wave propagation behavior of the three configurations is very similar. The open and closed configurations have very similar, but not identical, wave propagation behaviors. The behavior of the intermediate configuration shows a noticeably higher, albeit still small in absolute terms,  higher preference for energy propagation in the vertical direction than the other two configurations.  The horizontal direction is the preferred direction for energy flow in all three configurations because the changes in configuration do not significantly alter the alignment of the high longitudinal stiffness elements. So, they all behave like P-wave guides in the horizontal direction. All three configurations exhibit partial bangaps in the $X S$ and $Y \Gamma$ directions, and these band gaps have similar widths and central frequencies.  Hence, this PXCM can be effectively used as a low-pass mechanical filter along the vertical direction ($Y \Gamma$). 

The very small differences in the wave propagation behavior for the open and closed configurations of the unit cell imply that the wave propagation behavior of a PXCM that is in phase 1 (all cells are open) is very similar to that when the material is in phase 2 (all cells are closed).  It is not clear from this work if the small differences between the open (or closed) configurations and the intermediate configuration of the unit cell can be exploited to create a mechanical filter with programmable directionality of energy flow in the vertical direction.

\subsection{Transient analysis for a representative volume element}
In the previous two sections, we obtained the dispersion diagrams and directionality plots for the unit cell used in the PXCM studied here. These were obtained from Bloch analyses and hence are predicated on the periodicity of the material. Exploiting the periodicity of the PXCM in this manner significantly reduces the computational cost of simulating the wave propagation behavior in these materials. A representative volume element (RVE) is a 'slice' of an infinite periodic material sample that comprises sufficiently many unit cells so that it exhibits behavior that is representative of the material. In this section, we use direct integration transient finite element simulations to study the wave propagation behavior in an RVE of the PXCM.  This serves the following two purposes. Since this approach does not presume periodicity in the structure,  it provides a means to check the behavior obtained from the Bloch analyses of a unit cell. While the direct integration transient analyses are computationally very intensive even for an RVE, they provide a visualization of the wave propagation through the material. This provides valuable insight into the behavior of the material.

\begin{figure}[H]
\centering
\subfloat[(a)]{\includegraphics[height=10cm]{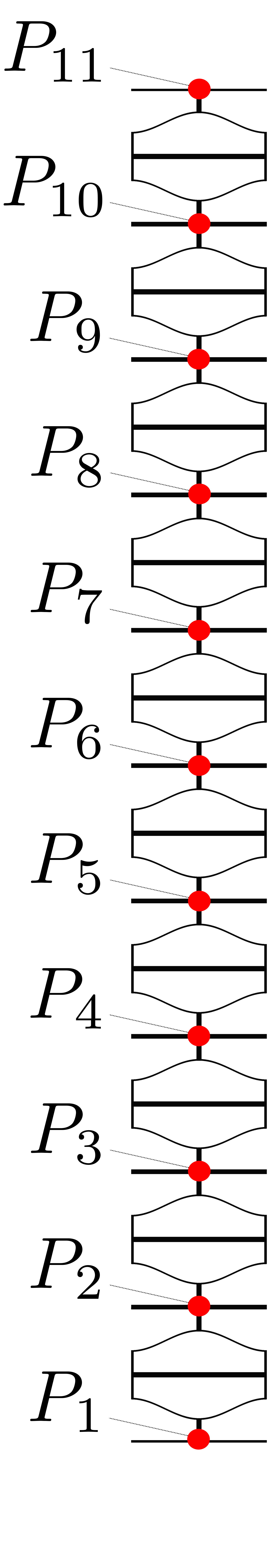}} \qquad
\subfloat[(b)]{\includegraphics[height=10cm]{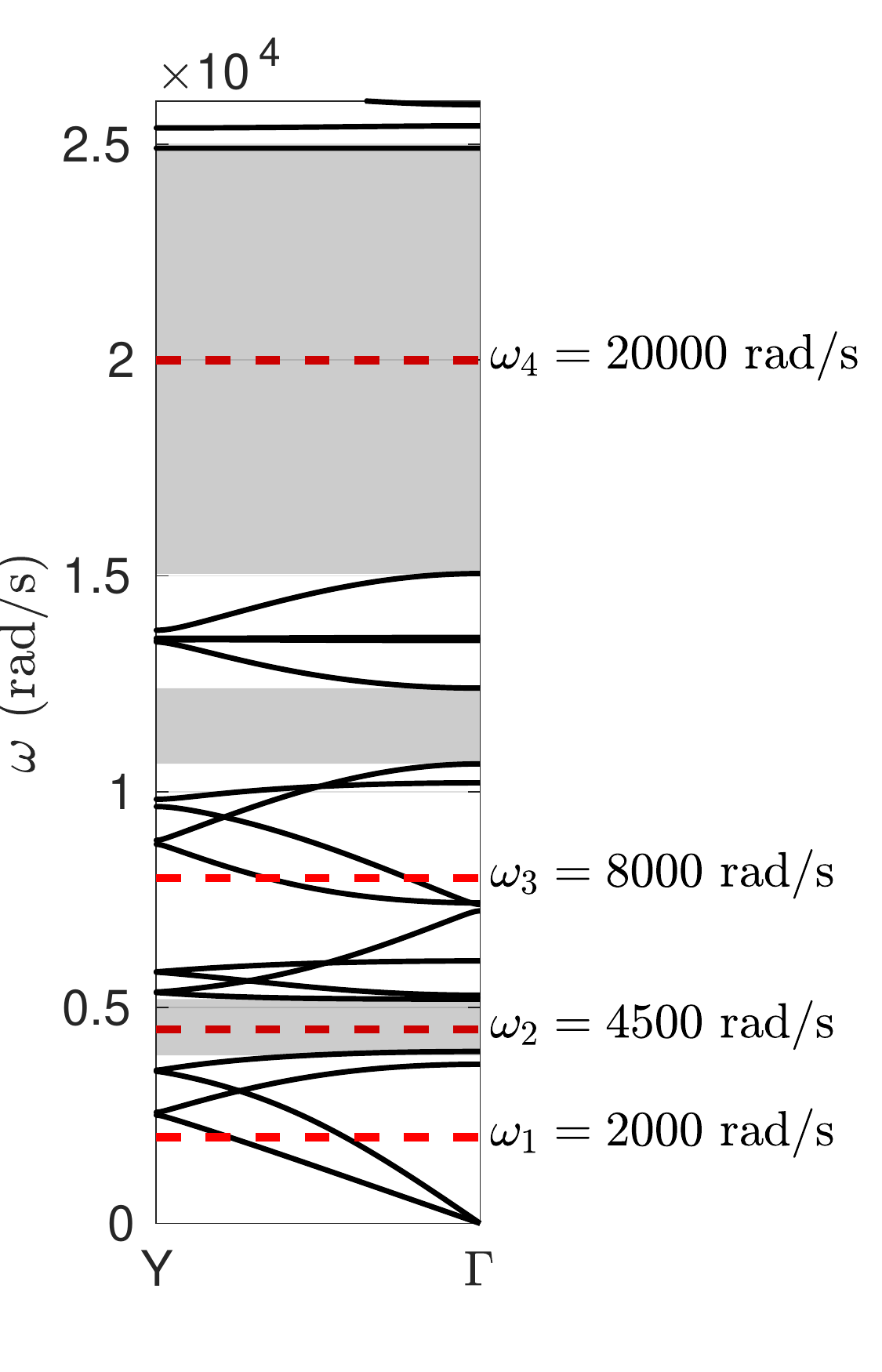}} \qquad
\subfloat[(c)]{\includegraphics[height=10cm]{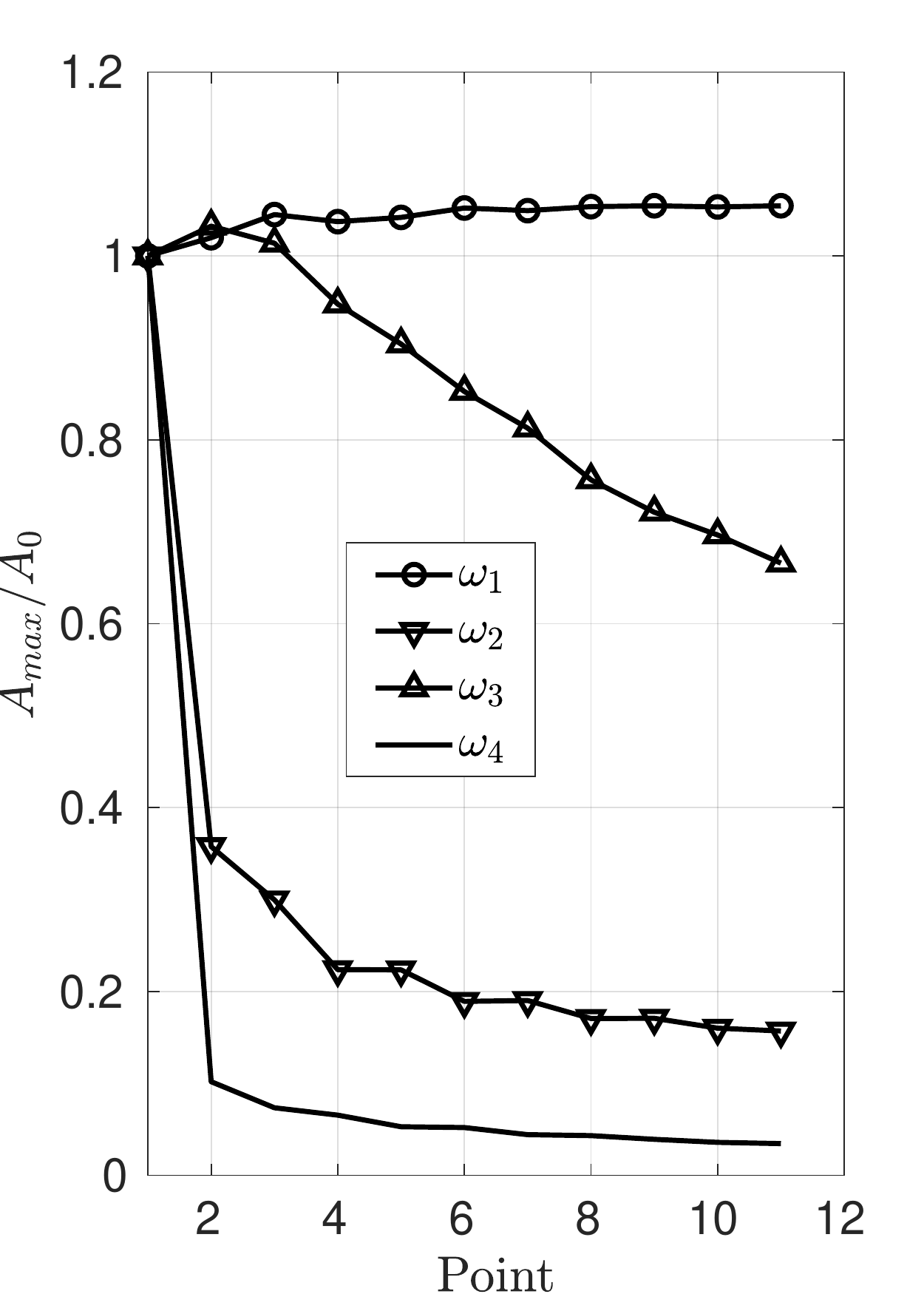}}
\caption{Selection of excitation frequencies for the transient analyses and displacement transmissibility along the length of the RVE. (a) RVE and selected points to measure displacement. (b) Dispersion diagram in the $Y \Gamma$ direction for open cell configuration. Red dashed lines indicate the chosen frequencies. (c) Displacement transmissibility for excitation frequencies $\omega_1$-$\omega_4$.}
\label{fig:freq}
\end{figure}

We focus attention on the response of the material to waves propagating in the vertical direction. Accordingly, a stack of ten unit cells in the vertical direction is chosen as the RVE as shown in \textbf{\cref{fig:freq}(a)}. All cells are in the open configuration. The RVE is excited at the bottom boundary by an imposed harmonic displacement (amplitude = 1.0 mm) which produces an incident P-wave traveling along the vertical direction ($Y \Gamma$). Roller supports are used on the sides of the RVE while an absorbing boundary is used for the top surface. This setup allows us to estimate the transient response of a much bigger sample of the PXCM.  As the displacements and strains are both expected to be small, we use linear finite element analyses for this study. 

We simulate the response of the RVE at four distinct frequencies that are chosen using the dispersion diagram for the unit cell in the open configuration (see \cref{fig:freq}(b)).  These frequencies are:  $\omega_1 = 2000$ rad/s, $\omega_2 = 4500$ rad/s, $\omega_3 = 8000$ rad/s and $\omega_4 = 20000$ rad/s.  Note that $\omega_2$ and $\omega_4$ are located inside two different stop bands (i.e. band gaps), whereas $\omega_1$ and $\omega_3$ are located in two different pass bands in the $Y \Gamma$ direction. 

From the time domain simulations, we measure how mush energy is transmited from one cell to another. The displacement transmissibility at a node $i$ is defined as the ratio ($X_i / X_0$), where $X_i$ and $X_0$ are the amplitudes of motion at node $i$ and the excitation boundary respectively. The displacement transmissibility is a quantitative measure of the attenuation produced by the PXCM material.  \Cref{fig:freq}(c) shows the displacement transmissibilities for the RVE comprising open cells at all four frequencies that were simulated in this section. The transmissibilities were computed  at all junctions between adjacent unit cells as shown in  \cref{fig:freq}(a). We observe that the displacement transmissibility decreases exponentially with distance from the source for $\omega_2$ and $\omega_4$, which lie inside band gaps for the material.  Near 60\% of the input energy remains trapped in the first cell for $\omega_2$ while most of the energy is transmitted to the end of the RVE for $\omega_1$ and $\omega_3$.  The material shows more extreme attenuation for  $\omega_4$ where 90\% of the energy remains localized in the first cell. We observed earlier that within the pass bands there were differences in the speed of propagation of the waves through the material.  The displacement transmissibilities for $\omega_1$ and $\omega_3$ show that there are corresponding differences in the attenuation of the energy as it propagates through the material at the two frequencies.  There is a small \emph{amplification} (transmissibility $>$ 1) for excitations applied at $\omega_1$ such that the amplitude motion of succesive junctions increases as we move away from the source. On the other hand, while the motion propagates all the way through the RVE for excitations at $\omega_3$, the motion is attenuated as we get farther away from the excitation source. 

Some snapshots and videos are included in the supplementary material. They show qualitative results from the time-domain simulations for open, semi-closed and closed configurations of the unit cell. From the videos, a close look at the deformation modes of the various structural elements of the unit cells in the RVE suggests that the central bar plays a fundamental role in the response of the RVE.  When the propagation is fully developed, the bar moves in phase with the sinusoidal beams for the frequencies in the pass bands, and it moves out of phase with the sinusoidal beams for frequencies in the band gaps.  The out of phase motions of the central bar and the sinusoidal beams give rise to destructive interference which hinders the onward propagation of energy.

\subsection{Modal analysis and tuning of the first band gap}
Transient simulations showed that the central bar moved out of phase with the sinusoidal segments for excitations in the two band gaps  that were explored earlier.  In this section, we investigate this connection further and use it to tune the band gaps in the $Y \Gamma$ direction for the PXCM material.

\begin{figure}[H]
\centering
\includegraphics[width=14cm]{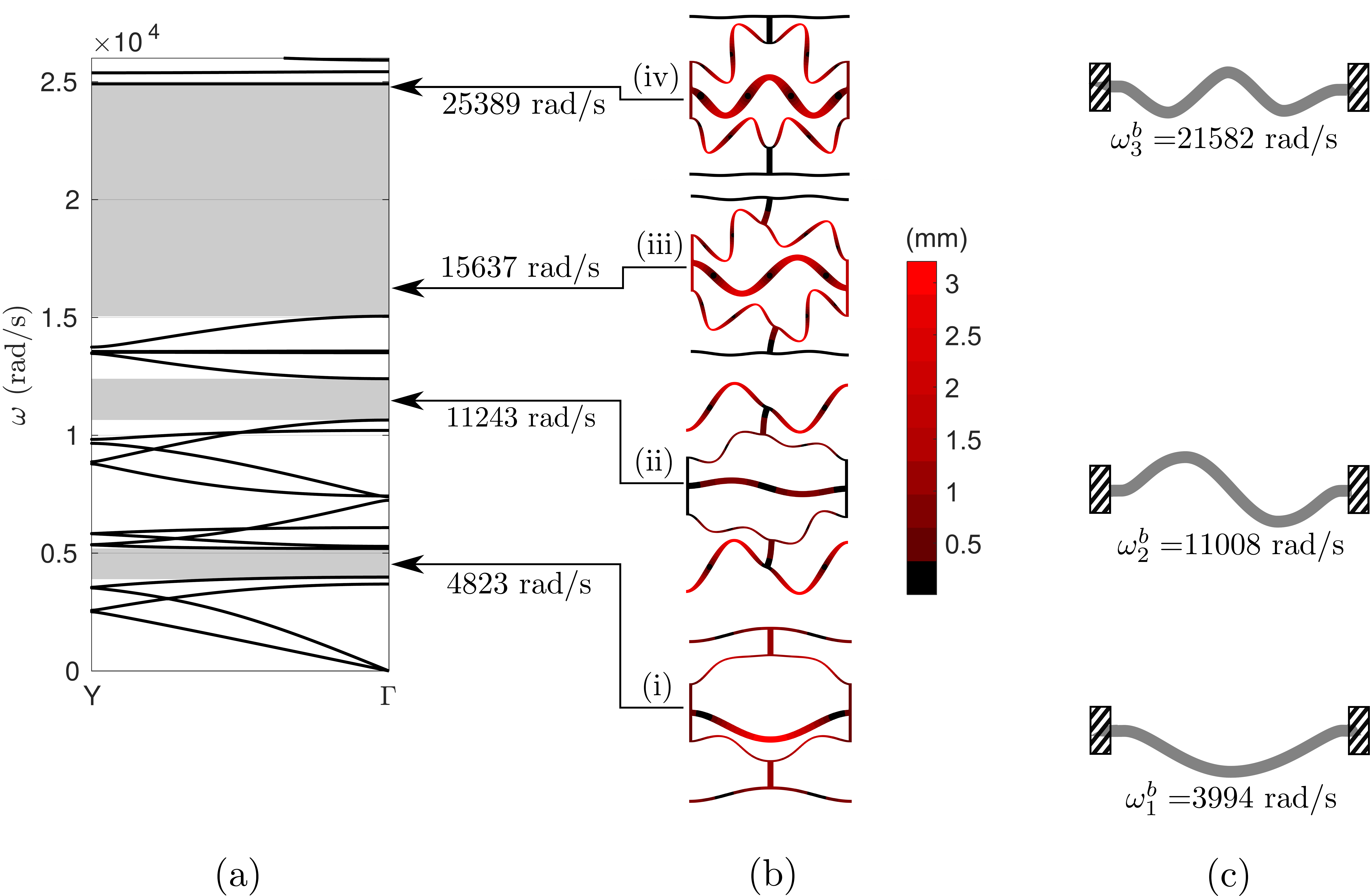}
\caption{Modal response of the unit cell in the open configuration. (a) Dispersion diagram in the $Y \Gamma$ direction for open cell configuration. (b) Modal shapes and their frequency location inside the dispersion diagram. (c) First three fundamental modes of a fixed-fixed beam. \label{fig:modal}}
\end{figure}

Salient mode shapes and frequencies from the modal analysis of a single unit cell in the open configuration are shown in column (b) of \textbf{\cref{fig:modal}}. The motion of the unit cell is unrestrained in this analysis except the left and right edges, which are constrained to move only in the vertical direction.  The modes shown here are chosen because they lie in the band gaps of the PXCM material for waves propagating in the vertical direction. We observe that the modes with frequencies $\omega_{u1}= 4823$ rad/s, $\omega_{u2}= 11243$ rad/s and $\omega_{u4}= 25389$ rad/s, all feature the central bar moving out of phase with respect to the sinusoidal beams.  Two of these frequencies ($\omega_{u1}= 4823$ rad/ and $\omega_{u4}= 25389$ rad/s) are close to the band gap excitation frequencies that were used in the transient RVE simulations ($\omega_{2}= 4500$ rad/s and $\omega_{4}= 20000$ rad/s respectively). Recall that when the unit cell is subject to a harmonic excitation at a frequency near a natural frequency, we expect the corresponding mode to contribute significantly to the overall motion of the cell. Thus,  we can establish a connection between the modes of a unit cell and band gaps in the $Y \Gamma$ direction for this PXCM. 

We can take this reasoning a step further. A closer look at the modes corresponding to $\omega_{u1}, \omega_{u2}$ and $\omega_{u4}$ shows that the deformation of the central bar in these modes of the unit cell is similar to that in the first three modes for a fixed-fixed beam in bending.  The natural frequencies and mode shapes for these modes are shown in column (c) of figure \cref{fig:modal}. The close correspondence between these frequencies and the band gaps in the $Y \Gamma$ direction for the PXCM suggests that these are \emph{locally resonant bandgaps} that arise when the frequency of excitation approaches the resonance frequencies of some structural elements  in the cell (\cite{celli2015manipulating,liu2000locally,sigalas1992elastic}). Such band gaps are associated with energy localization, which is consistent with our observations on the transient responses of the RVE in the previous section. We can now exploit the above correspondence to tune these band gaps.

\begin{figure}
\centering
\subfloat{\includegraphics[width=\textwidth]{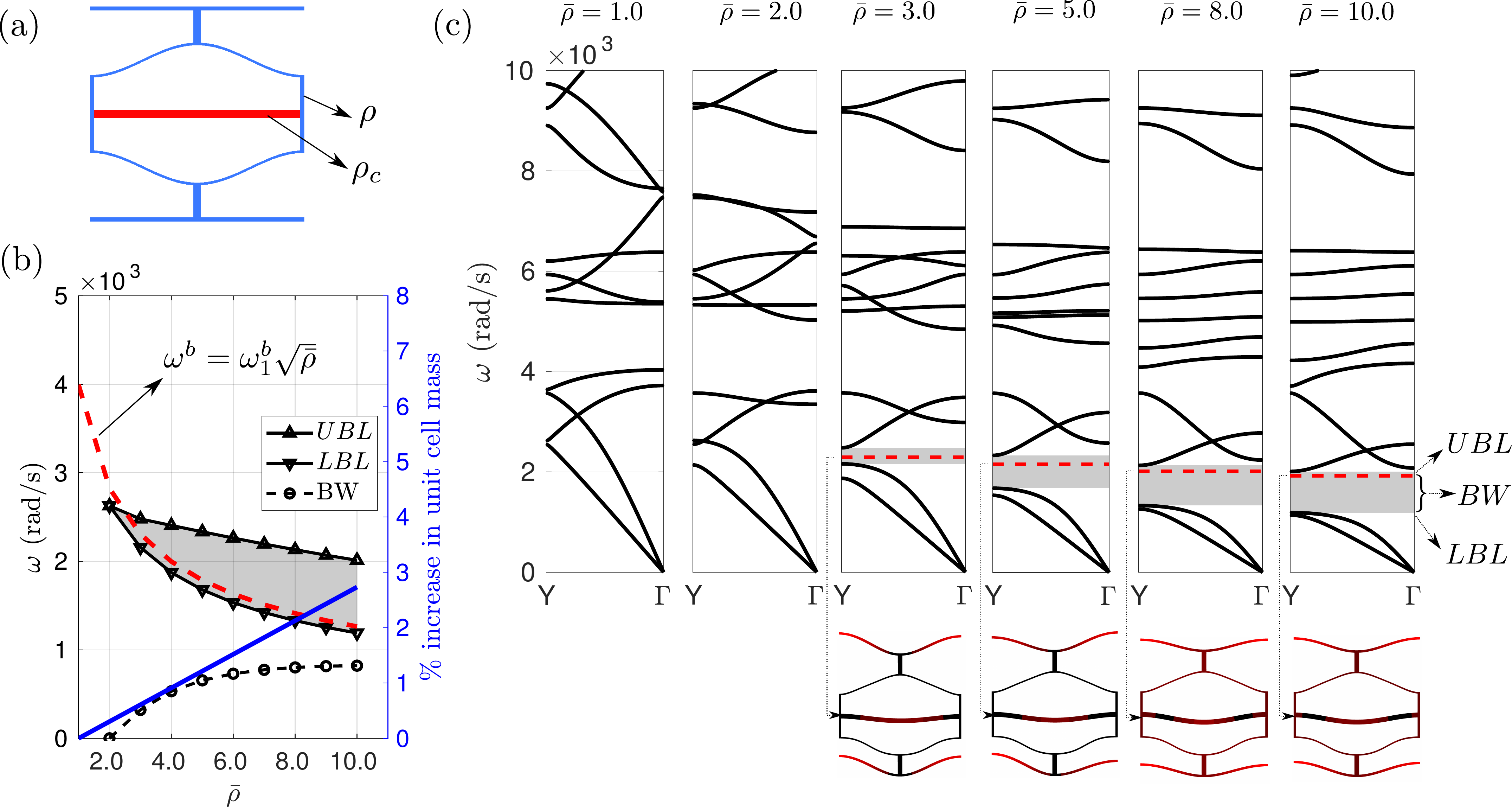}}

\caption{Modulating the partial band gap in the $Y \Gamma$ direction for the PXCM by varying the density of the material used for the central bar. (a) Unit cell geometry - blue region: density $\rho$, red region: density $\rho_c$. (b) Evolution of $UBL$, $LBL$ and $BW$ for the PXCM with increasing $\bar{\rho}$. (c)  Dispersion relations for various values of  $\bar{\rho} = \rho_c / \rho$. For values of $\bar{\rho}>2.0$, a modal shape within the new band gap and involving the central beam, was always found.}
\label{fig:incl}
\end{figure}

The dynamic behavior of the central bar can be modulated by varying its material properties (e.g. modulus and density) or its geometry (e.g. cross-section dimensions, non-prismatic profile, etc.). Closed form analytical expressions can be used to estimate the natural frequencies of the central bar for any of these modifications. Thus, we can readily determine the magnitude and nature of the modifications needed to achieve a target band gap frequency for the PXCM. In this study, we choose to modulate the density of the material used for the central bar while keeping all other attributes of the central bar and the rest of the unit cell unchanged.  This is indicated in \textbf{\cref{fig:incl}}(a) by using a different color for the central beam as compared to the rest of the unit cell. Let $\rho_c$ and $\rho$ denote the densities of the materials used for the central beam and the rest of the unit cell respectively.  We compute the dispersion diagrams for various values of the parameter $\bar{\rho} = \rho_c / \rho$ ratio in the range $[1.0 , 10.0]$.  \textbf{\Cref{fig:incl}}(c) shows the $Y \Gamma$ part of the dispersion diagram for salient values of this parameter. We note that a partial band gap in the $Y \Gamma$ direction does not exist around $\omega = 2000$ rad/s in the original ($\bar{\rho} = 1.0$) unit cell; the lowest frequency band gap  in the original cell is around  $\omega = 4000$ rad/s. A partial band gap around $\omega = 2000$ rad/s does not appear  until $\bar{\rho} > 2.0$. The band gap stretches from a lower limit of $LBL$ to an upper limit of $UBL$ (see \cref{fig:incl}(c)). The band gap width is defined as $BW = UBL - LBL$. Further increases in the value of $\bar{\rho}$ beyond $2.0$ lead to 1) a monotonic reduction in both $LBL$ and $UBL$, and 2) a concomittant monotonic increase in $BW$. This variation in the band gap frequency interval  (especially, the $LBL$) closely follows the dependence of the first natural frequency for a fixed-fixed beam in bending on the density of the beam material. This dependence ($ \omega^b = \omega^b_1 / \sqrt{\bar{\rho}}$) is indicated on figure \cref{fig:incl}(b) by a red dashed line.\\

The results in this section show that key insights gained from a step-wise deconstruction of the unit cell and transient analyses of an RVE enable us to tune the partial band gaps along the $Y \Gamma$ direction in the PXCM in a straightforward manner. Although, we illustrated this ability by tuning only the lowest band gap, it should be evident that the other resonance band gaps in the material's response can also be tuned in a similar manner.

\section{Conclusion}
We observed relatively minor differences in the dispersion diagrams and directionality of energy flow for the PXCM when all of its cells were in the open configuration versus when they were all in the closed configuration. This seems susprising given that the change in configuration associated with a significant change in the specific volume of the material, periodicity in the vertical direction and the shape of the sinusoidal beams in the unit cell. As discussed in Section 1, all of these changes have been shown to be associated with large changes in the wave propagation behavior by other researchers in the area.  This seeming contradiction can be resolved by a closer look at the specifics of this material.

The unit cell is symmetric with two individual bent beam mechanisms that are mirror images. Although each sinusoidal beam by itself undergoes a large change in shape when the unit cell transitions from the open configuration to the closed one, by simply transposing the top and bottom beams we can see that the propagation pathways available to an incident wave are substantially similar in both configurations.

The maximum strain in the unit cell when it is in the closed stable configuration is $2.5 \times 10^{-6}$.  Thus, the pre-stress is quite low and cannot, by itself, induce a noticeable change in the wave propagation behavior of the material.

Meaud and Che \cite{meaud2017tuning} have shown that a large change in the linear wave propagation behavior can be obtained in a one dimensional chain of bistable building blocks, that are similar to the curved beams that make up the unit cell of the PXCM, if the thickness of the beams is chosen to be near a threshold value that separates bistable unit cell behavior from monostable behavior.  Near this threshold value the ratio $k_u/k_d \rightarrow \infty$ because $k_d \rightarrow 0$, where $k_u$ and $k_d$ are the stiffnesses of the beam in its undeformed and deformed stable configuration respectively.  They note that  the dramatic change in wave propagation behavior near this threshold comes at the expense of stability of the unit cell as the cell can easily switch between the two stable configurations even for waves of low amplitude. Thus, their design is limited to waves of infinitesimal amplitude.

There are small but significant differences in the work reported in this reference and the current work. These arise mainly from the difference in the curved beam shape, and the periodicity of the boundary conditions for a two-dimensional material instead of a one dimensional one. However, despite these differences, their results are relevant to this discussion.  The unit cell design in this work has $Q= 12.2 >> 2.4$, which is the boundary between monostable and bistable unit cell behavior. The ratio of the stiffnesses of the unit cell in the open and closed stable configurations for loads applied in the vertical direction is 2.79. Therefore, the unit cell in this PXCM is far removed from the point in the design space where a significant change in the wave propagation behavior between the open and closed configurations was reported by Meaud and Che \cite{meaud2017tuning}.  As a further confirmation of this difference, when we rescale the dimensions of our unit cell to allow our design to be overlaid on the parameteric band gap maps reported in that reference, we observe that our design lies in the parts of the maps where there is little or no difference between the wave propagation behaviors of the open and closed cell configurations.

Meaud and Che \cite{meaud2017tuning} and Matlack et al. \cite{matlack2016composite} vary the stiffness of some elements in the unit cell to modulate the location and width of the band gaps.  The band gaps can be driven to lower frequencies, as is desirable, by decreasing the stiffness of the unit cell elements.  In the context of the unit cell design for the PXCM studied here, that would require the stiffness of the sinusoidal beams to be reduced by reducing their thickness. This is not desirable because it reduces the quasi-static energy dissipation capacity of the PXCM.  Therefore, we choose to tune the bandgap width by modulating the mass of the central horizontal beam instead of the thickness of the curved beams.  We showed that the band gap frequency can be lowered by increasing the mass of the central beam.  This allows us to tune the  linear wave propagation behavior of the PXCM without adversely affecting its energy dissipation behavior.

As discussed in the previous section, the partial band gap in the $Y \Gamma$ direction created by modulating the mass of the central beam is due to the local resonance of the central beam.  Resonant bandgaps are inherently narrow.  For a $\bar{\rho} = 10$, we observed a normalized band gap width of 0.5.
Following Krodel et al. \cite{krodel2015wide}, we can widen this band gap by extending the unit cell such that the enlarged unit cell comprises multiple current unit cells each of which has a central bar that is tuned to a different frequency in the interval of the desired band gap. 
The low frequency resonant band gap could also be widened by coupling it with a nearby Bragg scattering band gap \cite{yuan2013coupling}. In \cref{fig:incl}(b), we observe a Bragg band gap around 5000 rad/s. The resonance frequency for a fixed-fixed central bar at $\bar{\rho}= 1.0$ is approximately 4000 rad/s. We can increase the resonance frequency of the central beam until the associated resonance band gap gets coupled to the Bragg band gap by increasing the thickness of the central beam.  This would allows us to achieve a wide low frequency band gap without adversely affecting the quasi-static energy dissipation capacity of the PXCM.

A final comment concerns the significance of the quasi-static energy dissipation capacity of the PXCM material to its function as a phononic metamaterial for waves propagating in the $Y \Gamma$ direction. Recall from \cref{fig:prog7-f} that the dispersion and energy propagation directionality of the unit cell designs 8 and 9 are substantially similar for linear wave propagation. The latter design  exhibits bistability but the former does not.  Nadkarni et al. \cite{nadkarni2014dynamics} have noted that mechanical metamaterials based on negative stiffness elements can provide benefits in the nonlinear regime that are not seen in the linear  regime. Frazier and Kochmann \cite{frazier2017band} show that a one dimensional elastic chain with bistable resonators continues to attenuate wave propagation in a band gap even at high wave amplitudes unlike linear phononic metamaterials which allow wave propagation inside a band gap for high amplitude waves via a mechanism called 'supratransmission.' Although this study was limited to the linear wave propagation behavior of a PXCM, we anticipate that this PXCM would function effectively as a phononic metamaterial  for high amplitude waves because of its bistable unit cells. In contrast, a periodic cellular material based on the unit cell design 8 could function as a phononic metamaterial only for low amplitude waves.

While the results  reported in this work were based on the deconstruction of the wave propagation behavior of a particular PXCM, we believe that the underlying concepts and insights are more general and can be used for designing new PXCMs for phononic metamaterial applications.

\section{Acknowledgements}
This work was supported by EAFIT and COLCIENCIAS' Scholarship Program No. 6172; and by Fulbright under the Colombian Ph.D. Student program.  We would like to thank Sara E. Rodr\'{i}guez G\'{o}mez  for valuable technical discussions.  NDM would like to thank Nancy L. Johnson and Paul E. Krajewski of the Vehicle Systems Research Lab for their support and encouragement. N. Mankame, P. Zavattieri and D. Restrepo gratefully acknowledge the generous financial support of Nation Science Foundation through GOALI award CMMI-1538898.

\appendix
\section{Algorithm to calculate directionality behaviour}
\label{app:dir}

\begin{algorithm}[H]
 \KwData{Modes from Bloch analysis, $tol$, $n_{\theta}$}
 \KwResult{Final directional behaviour $D$}
 initialization of $D$, $e$, $i$;
 \textbf{Comment:} Arrays size: $D(n_{\theta})$, $d(n_{\theta})$\\
 \While{$e > tol$}{
  $V \leftarrow \nabla M_i/\alpha$ \;
  $G \leftarrow$ magnitude of $V$ (group velocity) \;
  \textbf{Comment:} Vector count over the vector field $V$: $d \leftarrow C(V)$\\
  initialization of $d$\;
  \For{ j $\leftarrow$ vectors in V}{
    dir$_{\theta}$ $\leftarrow$  calculate direction of vector $V_j$ and assignment inside one angular bin\;
    $d($dir$_{\theta}) \leftarrow d($dir$_{\theta})$ + $G_j$\;
    \textbf{Comment:} $d$ informs how much group velocity is present inside each angular section \\
    }
    \textbf{Comment:} Update of $D$\\
   $D \leftarrow D + d $\;
   \textbf{Comment:} Update of $e$ \\
   $e \leftarrow$ compute error\;
   $i \leftarrow i + 1 $\;
 }
 \caption{Detailed calculation process of directional behaviour}
 \label{alg:dir}
\end{algorithm}

\bibliographystyle{unsrt}
\bibliography{PXCMref} 


\end{document}